\documentclass[useAMS,usenatbib]{mn2e}
\usepackage{amssymb}
\usepackage{graphicx,epsfig}
\title{Correlation function of quasars in real and redshift space from Sloan Digital Sky Survey Data Release 7}
\author[G. Ivashchenko, V. I. Zhdanov, A. V. Tugay]{G. Ivashchenko$^{1}$\thanks{E-mail:
g.ivashchenko@gmail.com}, V. I. Zhdanov$^{1}$\thanks{E-mail: zhdanov@observ.univ.kiev.ua} and A. V. Tugay$^{2}$\thanks{E-mail: tugay@univ.kiev.ua}\\
$^{1}$ Astronomical Observatory of National Taras Shevchenko University of Kyiv, Observatorna str., 3, 04058, Kyiv, Ukraine \\ $^{2}$ Faculty of Physics of National Taras Shevchenko University of Kyiv, Glushkova ave., 2, 03127, Kyiv, Ukraine}
\begin{document}

\date{Accepted 2010 July 23. Received 2010 July 5; in original form 2010 May 16}

\pagerange{\pageref{firstpage}--\pageref{lastpage}} \pubyear{2010}

\maketitle

\label{firstpage}

\begin{abstract}
We analyze the quasar two-point correlation function (2pCF) within the redshift interval $0.8<z<2.2$ using a sample of 52303 quasars selected from the recent 7th Data Release of the Sloan Digital Sky Survey. Our approach to 2pCF uses a concept of locally Lorentz (Fermi) frame for determination of the distance between objects and  permutation method of the random catalogue generation. Assuming the spatially flat cosmological model with given $\Omega_{\Lambda}=0.726$, we found that the real-space 2pCF is fitted well with the power-low model within the distance range $1<\sigma<35$~$h^{-1}$~Mpc with the correlation length $r_{0}=5.85\pm0.33$~$h^{-1}$~Mpc and the slope $\gamma=1.87\pm0.07$. The redshift-space 2pCF is approximated with $s_{0}=6.43\pm0.63$~$h^{-1}$~Mpc and $\gamma=1.21\pm0.24$ for $1<s<10$~$h^{-1}$~Mpc, and $s_{0}=7.37\pm0.81$~$h^{-1}$~Mpc and $\gamma=1.90\pm0.24$ for $10<s<35$~$h^{-1}$~Mpc. For distances $s>10\,h^{-1}$~Mpc the parameter describing the large-scale infall to density inhomogeneities is $\beta=0.63\pm0.10$ with the linear bias $b=1.44\pm0.22$ that marginally (within 2$\sigma$) agrees with the linear theory of cosmological perturbations. We discuss possibilities to obtain a statistical estimate of the random component of quasars velocities (different from the large-scale infall). We note rather slight dependence of quasars velocity dispersion upon the 2pCF parameters in the region $r<2$~Mpc.
\end{abstract}


\section{Introduction}\label{sec:1}

\indent\indent Spatial distribution of quasars is one of the main sources of observational information about the largescale structure of the Universe; this is at present almost the only source of statistical information about matter inhomogeneity and clustering at cosmological redshifts. The most important results on quasars clustering were obtained using two largest quasar surveys up to date: the 2-degree Field Quasar Survey with 2QZ catalogue as a result (http://www.2dfquasar.org, \citet{Croom_1998}) and the Sloan Digital Sky Survey (SDSS; http://www.sdss.org), which has been completed with the 7th release \citep{Abazadjian_2009}. The two-point correlation functions  (2pCF) \citep{peebles_book} of quasars $\xi(r)$ are important characteristics of matter spatial inhomogeneity that may be compared to cosmological theories of structure formation. Development of 2dF and SDSS surveys gave a powerful incentive to investigation of various aspects of correlation functions of quasars (see, e.g., \citet{croom_2005,myers_2006} and  references therein). In the present paper we use the 7-th data release of SDSS to study 2pCF of quasars.

The main problems concerning  reconstruction of the cosmological mass distribution from redshift surveys are as follows. The first problem is that the surveys of extragalactic objects give us an information only about distribution of the luminous matter which is biased relative to the dark matter \citep{dekel_1987}. Biasing may depend on the physical peculiarities of extragalactic objects, e.g., on morphological type \citep{einasto_2007,ross_2007}, luminosity \citep{beisbart_2000,sorrentino_2006}, color-index \citep{coil_2007}, star formation rate \citep{owners_2007} etc. and evolves with redshift \citep{croom_2005,myers_2006,myers_2006_1,weinstein_2004,porciani_2004,daAngela_2008,Mountrichas_2009}. Commonly used supposition of linear biasing means that the density variations of certain kind of objects (in our case quasars) is proportional to that of the whole matter; therefore
\begin{equation}\label{bias1}
\xi(r)=b^2\,\xi_{m}(r),
\end{equation}
where  $\xi_{m}(r)$ is 2pCF of the whole matter, $\xi(r)$ is 2pCF of quasars and $b$ is the bias parameter.

The second problem of the  3-dimensional analysis of the matter distribution is due to the fact that the observed redshifts of
extragalactic objects are `contaminated' by measurement errors and non-Hubble motions. The distances calculated from these redshifts
without taking into account the unknown peculiar velocities are called distances in redshift-space (in contrast to the real space). As our Universe is isotropic, the correlation function must be spherically symmetric in the real-space. But in the redshift-space it appears to be distorted. On  smaller scales the profile of galaxies 2pCF is stretched along the line of sight (`Finger of God' effect) due to virial velocities of objects inside the galaxy clusters (this can be neglected for quasars if we consider $z>1$), their random velocities and redshift errors. This effect is especially noticeable for quasar pairs with the projected linear separations $\lesssim2$~Mpc, where the SDSS data are expected to be incomplete \citep{hennawi_06}. On larger scales the 2pCF profile is flattened along the line of sight (`The Bull's eye' or \citet{kaiser_1987} effect) due to the gravitational infall to density inhomogeneities; this kind of redshift-space distortion dominates on the linear scales. These effects are parameterized by line-of-sight pairwise velocity dispersion $\langle w^{2}\rangle^{1/2}$ and infall parameter $\beta$. It is worth to note that the non-Hubble motions and redshift errors are not the only sources of redshift-space distortions. One more effect of geometric flattening that can lead to distortion of 2pCF can be due to a wrong choice of cosmological parameters $\Omega_{M}$, $\Omega_{\Lambda}$; this provides an additional tool for estimation of these parameters by means of a geometrical  test of \citet{alcock_paczynski_1979}. The redshift-space 2pCF of quasars can be in principle used for estimation of all parameters ($\beta$, $\langle w^{2}\rangle^{1/2}$ and cosmological ones) simultaneously. But due to a degeneracy between the geometric distortions and the redshift-space distortions (see, e.g., \citet{hoyle_2002}) this problem is complicated. \citet{hoyle_2002} and \citet{daAngela_2005} proposed a method that allows to break this degeneracy  by means of combination of the \citet{alcock_paczynski_1979} test with that based on evolution of the quasar clustering amplitude, which has a different dependence on $\beta(\bar{z})$ and $\Omega_{M}(0)$.

As we cannot estimate the non-Hubble motion of each quasar independently, the effects of the redshift-space distortion are at present the only source of statistical information about proper velocities of quasars. On the other hand, these effects prevent direct determination of 3D 2pCF that involves line-of-sight distances between objects determined from $z$-measurements. This urges us to use only the projected distances $\sigma$ (i.e. orthogonal to the line of sight, which may be considered as independent of proper motions) to determine the projected 2pCF and then to restore the real-space 2pCF. It is well known that such reconstruction of the real-space 2pCF from projected one is mathematically ill-posed problem. However, most authors avoid this difficulty using a concrete functional form of 2pCF. Typically 2pCF is represented in a power-law form $\xi(r)=\left(r_0/r\right)^{\gamma}$, though it is clear that 2pCF slope and correlation length may be different on different interval. For example, \citet{daAngela_2005} showed that double power-low model gives a good fit to 2pCF of quasars distribution with $\xi(r)=\left(6.0/r\right)^{1.45}$ over scales of $1<r<10$~$h^{-1}$~Mpc and $\xi=\left(7.25/r\right)^{2.30}$ over scales of $10<r<40$~$h^{-1}$~Mpc for a quasar sample from 2QZ catalogue with redshifts within the range $0.3<z<2.2$. Using quasar catalogue of spectroscopically confirmed quasars \citep{Schneider_2007} based on the 5th data release of SDSS \citet{ross_2009} obtained $r_{0}=5.45^{+0.35}_{-0.45}$~$h^{-1}$~Mpc and $\gamma=1.90^{+0.04}_{-0.03}$ over scales of $1<\sigma<130$~$h^{-1}$~Mpc  for low-redshift ($0.3<z<2.2$) quasars, and for high-redshift ($2.9<z<5.4$) quasars from the same catalogue \citet{shen_2007} obtained $r_{0}=15.2\pm2.7$~$h^{-1}$~Mpc and $\gamma=2.0\pm0.3$ over scales $4<\sigma<150$~$h^{-1}$~Mpc. Note that similar large correlation length was obtained for radio-sources by \citet{blake_2003} and for X-ray sources by \citet{plionis_2004} and \citet{basilakos_2004}; at least a part of both of these object types are considered to be AGNs. In addition to redshift dependence of the quasar clustering confirmed by \citet{croom_2005,myers_2006,myers_2006_1,porciani_2004,porciani_2006,daAngela_2008}, some authors \citep{porciani_2006,daAngela_2008,Mountrichas_2009} argue that quasar clustering depends weakly on luminosity. Hence speaking about representation of 2pCF in power-law form in the present paper we always mean that 2pCF is averaged over quasars with different luminosities and over certain redshift range and distance intervals described below. In particular, when dealing with the bias estimates we confine ourselves to the intervals where our results may be compared with the linear theory of matter inhomogeneity.

The most complete full investigation on non-Hubble motions of quasars together with cosmological parameters was done by 2dF team for 2dF QSO Survey (see \citet{croom_2005,daAngela_2005,hoyle_2002,outram_2004}). E.g. \citet{daAngela_2005} obtained $\Omega_{M}=0.35^{+0.19}_{-0.13}$ and $\beta=0.50^{+0.13}_{-0.15}$ (for the sample mean redshift $\bar{z}=1.4$). In these studies the value of pairwise velocity dispersion of quasars was usually treated as fixed parameter and its value was chosen either from Hubble volume simulations or following previous results on galaxy pairwise velocity dispersion assuming that this value does not evolve strongly with time. For galaxies, we have more definite results for the velocity dispersion obtained from the redshift-space distortions. The first estimate of \citet{davis_peebles_1983},  $340\pm40$~km/s, remained a canonical one almost a decade. However, later \citet{mo_1993} showed that the value of pairwise peculiar velocity dispersion depends strongly on the sample. Later this result was confirmed by values $540\pm180$~km/s, given by \citet{marzke_1995} for the Second Center for Astrophysics Galaxy Redshift Survey (CfA2) and the Southern Sky Redshift Survey (SSRS2), and $416\pm36$~km/s, given by \citet{ratcliffe_IV_1998} for Durham/United Kingdom Schmidt Telescope (UKST) Galaxy Redshift Survey and others. \citet{ratcliffe_IV_1998} also presented the value $\beta=0.48\pm0.11$, which agrees well with later results, such as the results of \citet{hawkins_2003} for 2dF Galaxy Redshift Survey. A similar value, 330~km/s, was obtained for luminous red galaxies (LRG) from the 2dF-SDSS LRG and QSO (2SLAQ) survey by \citet{ross_2008}.

In this paper, we present new results on the 2pCF  slope and correlation length, infall parameter $\beta$ using the last 7-th Data Release of SDSS. In Section~\ref{sec:2} we describe the quasar samples and technique for random catalogue generation we used in our study. In Sections~\ref{sec:3} and \ref{sec:4} we present the results on parameters of the real-space and redshift-space 2pCFs and techniques for their estimation. Section~\ref{sec:4} also contains estimations of the infall parameter $\beta$ and quasar bias $b$ from correlation function analysis on linear scales larger than $~10$~Mpc. In Section~\ref{sec:6} we discuss possibilities to use `Finger of God' effect to estimate the random quasars velocities, which does not correlate with density inhomogeneities. Because determination of random pairwise velocity dispersion $\langle w^2 \rangle^{1/2}$ of quasars needs knowledge of 2pCF in the region $r<2$~Mpc, which is not well studied, we considered different simplified models of 2pCF to study their outcome for $\langle w^2\rangle^{1/2}$. Finally, in Section~\ref{sec:7} we sum up the results.

Throughout the paper we use the spatially flat cosmological model with the combined estimates $\Omega_{\Lambda}=0.726$ and $h=0.705$
\citep{Komatsu_2009}, which agrees with recent estimate from seven-year \textit{Wilkinson Microwave Anisotropy Probe} (WMAP) data \citep{Larson_2010}.

\section[]{Data}\label{sec:2}

\subsection{The sample}\label{sec:2.1}

\indent\indent  Our sample is taken from the 7th Data Release of the Sloan Digital Sky Survey \citep{Abazadjian_2009}, containing about 100,000 quasars. The redshift range used in our analysis is $0.8\leq z\leq2.2$ and as the sky coverage of SDSS contains a one big piece and three narrow near-equatorial `stripes', we excluded these stripes to reduce boundary effects. As the data was taken from the list of all objects classified as quasars, which is not a vetted catalogue, we firstly excluded `bad' objects: objects with failed astrometry and photometry data. The second step was to
exclude objects with unreliable redshifts. The redshifts in SDSS are measured mainly using two techniques, emission-line measurements and cross-correlation, either both or one of these. We excluded objects with the redshift status `INCONSISTENT' and `FAILED'. The first status means that the redshift was measured by two techniques but the results were inconsistent. The second one means that the redshift measurement failed. The objects with redshift error $>0.01$ were also excluded.

\begin{figure}
\centering
\epsfig{figure=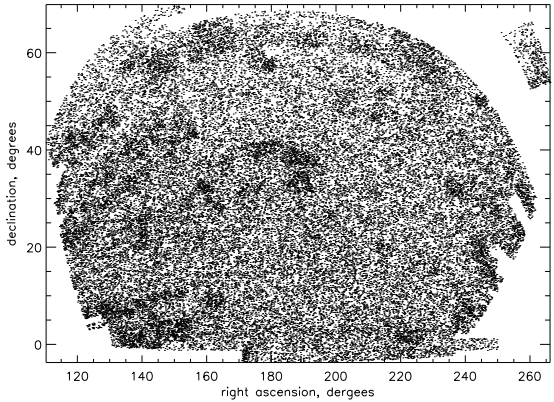,width=8cm}
\caption{Sky coverage of our \textit{full} quasar sample in equatorial coordinates.}
 \label{fig_1}
\end{figure}
\begin{figure}\centering
\epsfig{figure=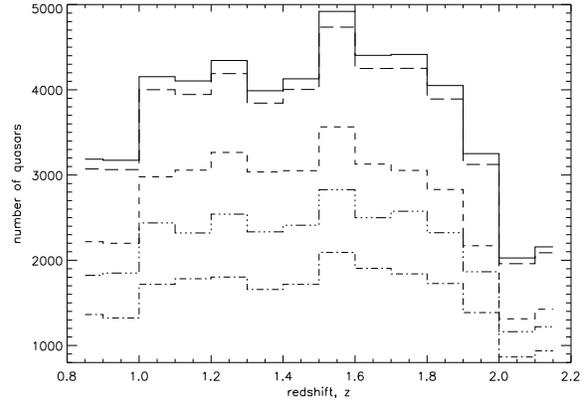,width=8cm}
\caption{Redshift distribution of our quasar samples: \textit{full} (solid line), \textit{uniform} (dashed), \textit{low-reddening} (dash-dot), \textit{high-reddening} (dash-dot-dot-dot), \textit{good} (long dashes)}
 \label{fig_2}
\end{figure}
\begin{figure}\centering
\epsfig{figure=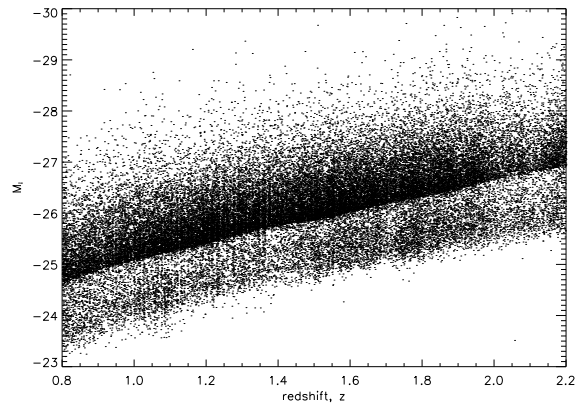,width=8cm}
\caption{Luminosity $-$ redshift distribution for \textit{uniform} sample.}
 \label{fig_2a}
\end{figure}

The third step was examination of all objects left after the previous steps by eye (photometry and spectroscopy data) using the SDSS data base. During these examination 466 objects were excluded as the ones having too faint and noisy spectra, or being fault `double' objects (i.e. single objects processed twice and thus appeared twice in the list with similar coordinates), or being artifacts of observations, or being stars (we have found 5 stars in this sample).

Our final sample, which we call \textit{full}, contains 52303 objects and has the  mean redshift $\bar{z}=1.47$. The sky coverage of the sample and its redshift distribution are shown in Figs.~\ref{fig_1} and \ref{fig_2} (solid line). The redshift-luminosity distribution of this sample is shown in
Fig.~\ref{fig_2a}, where the effect of $m_{i}=19.1$ magnitude limit can be seen. Here the values of absolute magnitudes in $i$-band, $M_{i}$, are calculated using $K$-correction from \citet{richards_2006}.

\subsection{Sample inhomogeneity}\label{sec:2.2}

\indent\indent As one can see from Fig.~\ref{fig_1} the sky coverage by our sample in not heterogeneous: some parts seem to be significantly denser then others. This is due the fact that these parts were twice spectroscopically covered and can influence our results. First of all we note that the algorithms we use to estimate the 2pCF parameters (sections 3.1 and 4.1) are just worked out in order to reduce the selection effects due to unphysical inhomogeneity \citep{ivzh_10,zhdan_ivashch2008}.

\begin{figure}
\centering
\epsfig{figure=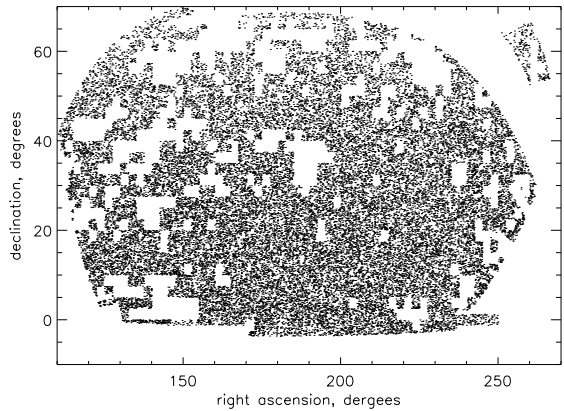,width=8cm}
\caption{Sky coverage of the \textit{full} sample with excluded patches with density higher than 0.5rms.}
 \label{fig_1a}
\end{figure}

Nevertheless, to test the influence of inhomogeneity we constructed 50 \textit{homogeneous} samples in the following way. The whole sky area covered by our sample was divided into $2.5^{\circ}\times2.5^{\circ}$ patches and the density of each patch, mean density ($\bar{n}=7.25$ objects per square degree) and its rms ($\sigma_{n}=2.55$ objects per square degree) were calculated. All the objects from the patches with density exceeding the mean one more than 0.5rms were excluded (see Fig.~\ref{fig_1a}). Then we generated 50 new \textit{homogeneous} samples by putting the excluded objects in their places, but choosing (in a random way) only those making the density of the given patch equal to the mean density. The sky coveredge of one of these samples is presented in Fig.~\ref{fig_1b}. Each new test sample has 35643 objects and a density $\bar{n}=6.35\pm1.19$ objects per deg$^{2}$. Apparent decrease of the density with increasing declination is the result of projection of the celestial sphere onto the plane.

\subsection{Tilable targets}\label{sec:2.3}

\begin{figure}\centering
\epsfig{figure=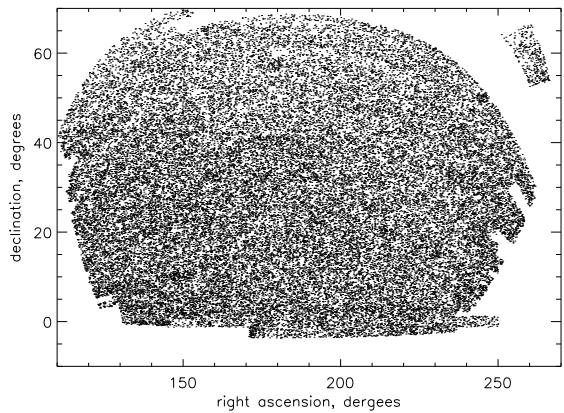,width=8cm}
\caption{Sky coverage of one of the \textit{homogeneous} samples.}
 \label{fig_1b}
\end{figure}

\begin{figure}\centering
\epsfig{figure=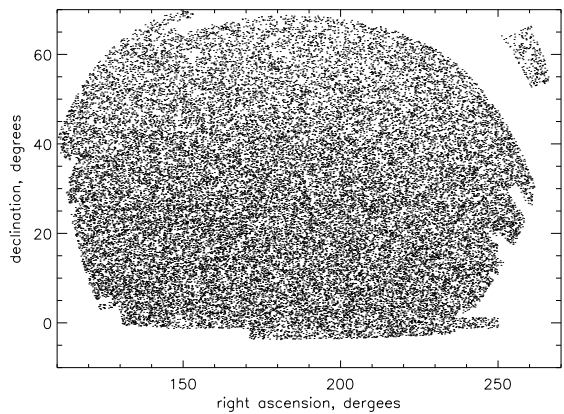,width=8cm}
\caption{Sky coverage of the \textit{uniform} sample.}
 \label{fig_1c}
\end{figure}

\indent\indent Most quasar candidates in the SDSS are selected based on their locations in multidimentional SDSS colour space and their cross-identification with radio sources from the Faint Images of the Radio Sky at Twenty-cm (FIRST) survey \citep{richards_2002}. Supplementing this primary quasar sample are quasars targeted by the GALAXY, X-RAY, STAR and SERENDIPITY SDSS software packages; no attempt at completeness was made for the last three categories. Thus, the sample we use is not complete and the possible effect of this incompleteness on our results has to be investigated. For this purpose we constructed the sample of the so-called `tilable' objects (\textit{uniform}) from our sample. According to SDSS glossary web-page, tilable targets are supposed to have as closed to uniform completeness as possible. These are targets with primTarget flag `QSO\_HIZ', `QSO\_CAP', `QSO\_SKIRT', `QSO\_FIRST\_CAP', `QSO\_FIRST\_SKIRT', `GALAXY\_RED', `GALAXY', `GALAXY\_BIG', `GALAXY\_BRIGHT\_CORE', `STAR\_BROWN\_DWARF', and second target flag `HOT\_STD'. The number of objects in the subsample constructed following this criterion is 37290. One can see that the sky coverage by the \textit{uniform} sample presented in Fig.~\ref{fig_1c} is close to uniform and close to the sky coverage by our \textit{homogeneous} samples (Fig.~\ref{fig_1b}). The redshift distribution of this \textit{uniform} sample is presented in Fig.~\ref{fig_2} (dashed line).


\subsection{Reddening}\label{sec:2.4}

\indent\indent As the main criterion of quasar candidates selection in SDSS is based on their magnitudes and colors, that have been corrected for Galactic extinction using the maps of \citet{schlegel_1998}, any systematic errors in the reddening model can induce additional effects on clustering results (see \citet{ross_2009} for discussion of possible effects). That is why following \citet{ross_2009} we devided our sample into low reddening ($0.0028<E(B-V)\leq0.0217$) sample and high reddening ($0.0127<E(B-V)\leq0.2603$) one. Values of $E(B-V)$ were obtained from the reddening in each SDSS band given in the catalogue and using the relative extinction for different bandpasses from Table~6 in \citet{schlegel_1998}. The numbers of objects in subsamples are 30188 and 22115 for low and high reddening correspondingly. The redshift distributions of these subsamples are presented in Fig.~\ref{fig_2} (dash-dot for \textit{low-reddening} sample and dash-dot-dot-dot for \textit{high-reddening} one).

\subsection{Bad fields}\label{sec:2.5}

\indent\indent Since the quasar selection algorithm involves colours and magnitude all photometric errors caused by e.g. photometric callibration can induce errors to this selection and result in additional inhomogeneity of the sample. For verification of these effect we constructed a \textit{good} sample using information about the data quality for each field (an image in all five bands with approximate dimentions of $13'\times10'$) from a field quality table, called RunQA. To construct the \textit{good} sample we used the FieldQAll number which describes the overall quality of a field and rejected the objects with its value equal to `bad'. This overall quality determination, FieldQAll, is based on the (dereddened) principal colors of all the stars in the field vs. the Galactic value of this quantity, the PSF quality, the difference between Aperture and PSF magnitudes for the same stars in the field, and the seeing. We also excluded the objects with rerun number equal to zero, as there are no fields with such rerun number in FieldQAll table. The number of objects in the \textit{good} sample is 50420. Its redshift distribution is presented in Fig.~\ref{fig_2} (long dashes).

\subsection{Fibre collisions}\label{sec:2.6}

\indent\indent  One another peculiarity of our sample is related to the so called `fibre collisions' effect, which is inherent for multifibre spectrographs. It means that two spectroscopic targets on the same plate cannot be measured simultaneously if they are closer than some distance ($55''$ for SDSS); it is only possible to obtain spectra for both such objects if they are covered by more than one plate \citep{Schneider_2007}. Angular separation $55''$ corresponds to $0.7-1.5$~Mpc projected distance for our redshift interval $0.8-2.2$. That is why we avoided this distance range when calculating the parameters of 2pCF. Of course this effect affects mostly the precision of our random velocities dispersion estimations due to the small sample of pairs with projected separation $<2$~Mpc. But in this case we cannot follow e.~g. \citet{ross_2009} supplementing their sample with quasar candidates from SDSS Quasar photometric catalogue with photometric redshifts \citep{richards_2009} due to the fact that the errors of photometric redshifts are about $\Delta z=0.3(2\sigma)$ which is much larger than the restrictions we put on our sample.

\subsection{Random samples}\label{sec:2.7}

\indent\indent An important point in reconstruction of the correlation function of extragalactic objects is generation of the random sample, which should reproduce a random distribution of objects with the same redshifts and angular distributions inherent to the initial sample as much as possible. For this purpose an imaging mask may be used (see e.g. \citet{myers_2006}). In the present work we used  the permutation method of random catalogue generation (see e.g. \citet{osmer_1981,zhdanov_2001,croom_2005}) that does not require any technical information about the survey. This method uses the random permutation of objects' redshifts from the initial list of quasars while keeping their coordinates unchanged. As a result one obtains a randomized list of objects with the same redshift distribution and the same distribution over the sky as in the initial sample. At the same time this randomization destroys correlations between objects of the initial sample: those objects which were clustered in the initial sample appear to be not clustered in the randomized sample as they have most probably different redshifts and thus are located far from each other.

\section[]{Real-space correlation function}\label{sec:3}

\subsection{The technique}\label{sec:3.1}

\indent\indent As peculiar velocities of objects distort the distance between objects along the line of sight, the parameters of the real-space 2pCF of quasars can be reconstructed from the projected 2pCF.  We use this standard approach, but taking into account some specific features of the permutation method of random catalogue generation.

2pCF $\xi(r)$ determines the probability $dP_i$ to find a neighbour of the object $Q_i$ from the sample at comoving distance $r$ in a volume element $dV$ as \citep{peebles_book}
\begin{equation}\label{2cf_definition}
dP_{i} = n_{i}[1+\xi(r)]dV,
\end{equation}
\noindent where $n_{i}$ is a number density of $i$-th object in the neighbourhood of $Q_i$. Here we must specify the definition of the distance between neighbours, which can be defined differently in curved space-time. For the distance scales under consideration this may be done with a sufficient accuracy using local Lorentz frames \citep{misner_etal}. For definiteness we use the Fermi reference frames and normal Fermi coordinates (Synge 1960; Misner et al. 1973); this is especially convenient because in case of spatially flat cosmological background this leads to distance relations analogous to that of Euclidean space (see Appendix).

We suppose that the real-space 2pCF of quasars has power-low form $\xi(r)=(r_{0}/r)^{\gamma}$ for distances $r<L$ and $r>L_0$, where $L$ is some characteristic comoving scale where 2pCF approaches to zero. We also make the following assumptions. First,  we neglect variations of $n_{i}$ on the scales $<L$. In fact, we also considered models that take into account variations of $n_{i}$ with redshift on this scale and found the resulting difference to be not statistically significant, as it can be expected from very beginning. Second, we assume, as usual, that quasars distribution is locally isotropic at any point of the real-space, i.e. it depends only on the distance between the objects. Third, we neglect variations of redshift distribution of the sample in different sky regions.

Further we pass to projected (comoving) separations. The only thing that makes our consideration different from the standard one \citep{peebles_book} is related to our method of the random catalogue generation. When counting the number of pairs with  projected separations within the range $[\sigma,\sigma+d\sigma]$, we include only pairs with line-of-sight separations $\Pi$ that satisfy the condition $\Pi<L$ \citep{zhdanov_2001, zhdan_ivashch2008}. (Here we denote the distance along the line of sight as $\Pi$ to distinguish it from the number $\pi$.) This number of pairs should be compared to analogous number in the randomized sample. If, e.g., we do not restrict $\Pi$ and include all pairs with the above projected separation, there obviously would be no difference in these numbers between the initial sample and the randomized one obtained by permutation of redshifts. On the other hand, we expect that almost all the pairs that have a noticeable correlation are concentrated within the region $\Pi<L$. Note that though at this point we use the observational values of $\Pi$ that involve contribution of the proper velocities, corresponding correction due to velocities here is negligible, because it affects only larger distances $\sim L$.

The total number of neighbouring quasars with the projected distances from a given quasar within the range $[\sigma,\sigma+d\sigma]$ and line-of-sight separations $\Pi<L$ is obtained by summing over all quasars from the sample
\begin{equation}
\label{DDsigma}
  DD(\sigma,d\sigma)\approx\sum\limits_{i} n_{i}\left[2L+
  \int\limits_{-L}^{L}\xi\left(\sqrt{\sigma^{2}+\Pi^{2}}\right)d\Pi\right]dS,
\end{equation}
where the sum is performed over all objects in the sample, $dS=2\pi\sigma d\sigma$. This number is twice larger than the number of pairs with separation from the above range, if the quasars and the neighbours are chosen from the same sample. In order to study boundary effects we also considered the case when the quasars are chosen from a subsample obtained from the initial sample with the exception of objects near the boundary, and the neighbours being looked for in the whole initial sample.

In case of the randomized redshifts the corresponding number of neighbours is
\[
 RR(\sigma,d\sigma)=2L\sum\limits_{i}n'_{i}dS,
\]
where $n'_{i}$ is the number density  in the neighbourhood of $i$-th object in the randomized sample.  As the redshift distribution of objects in initial and random samples are the same, than  one estimates $\sum\limits_{i}n_{i}\approx\sum\limits_{i}n'_{i}$ with a relative error $\sim1/\sqrt{N}$, where $N$ is the number of objects in the sample. Therefore
\begin{equation}
\label{DDRR}
 \frac{DD(\sigma,d\sigma)}{RR(\sigma,d\sigma)}\approx1+\frac{1}{2L}\int\limits_{-\infty}^{+\infty}\xi\left(\sqrt{\sigma^{2}+ \Pi^{2}}\right)d\Pi,
\end{equation}
where we extended the integration limits to infinity.

Note that possible difference of the ratio $\sum\limits_{i}n_{i}/\sum\limits_{i}n'_{i}$ from unity may be due to  variations of $z$-distribution of quasars in the different sky regions of the real survey. Corresponding correction was analyzed and found to be statistically insignificant \citep{zhdan_ivashch2008}.

In fact we used  not exactly Eq.~(\ref{DDRR}) but corresponding relations for separations in finite intervals $[\sigma, \sigma+\Delta\sigma]$; typically we used $\Delta\sigma=1$~Mpc. Then using the power-low form of 2pCF we have the following relations for fitting parameters $\gamma$ and $r_{0}$ (cf., e.g., \citet{shanks_1980,collins_1989}):
\begin{equation}\label{r_2pCF}
 \frac{DD(\sigma,\Delta\sigma)}{RR(\sigma,\Delta\sigma)}-1=B\frac{(\sigma+\Delta \sigma)^{3-\gamma}-(\sigma)^{3-\gamma}}{(3-\gamma)(\sigma +\Delta\sigma/2)},
\end{equation}
\noindent where
\[
B=\frac{\sqrt{\pi}\,\Gamma\left(\frac{\gamma-1}{2}\right)}{2L\,\Gamma(\frac{\gamma}{2})}r_{0}^{\gamma}.
\]

The errors of all the parameters in the present study have been estimated using the `jackknife' method (see, e.g., \citet{miller} for review), which has been successfully applied to clustering problems in extragalactic astronomy (see, e.g., \citet{myers_2006}). We divided our initial sample into $N_s\sim50-100$ equal stripes along declination; owing to large number of objects in each stripe, the stripes may be viewed as practically independent ones (here we neglect the boundary effects). Then new $N_s$ subsamples have been formed: $i$-th subsample is unification of all the stripes except $i$-th stripe. Jackknife dispersion estimate of any parameter $x$ calculated in this treatment is then
\begin{equation}\label{ftf_error}
 \sigma_{x}^{2}=\frac{N-1}{N}\sum\limits_{i=1}^{N}[x_{i}-\bar x]^{2},
\end{equation}
where $x_{i}$ is the corresponding value of this parameter for $i$-th subsample and $\bar{x}$ is either the value of $x$ for the initial sample or it is average over $N_s$ subsamples.
\begin{figure}
\centering
\epsfig{figure=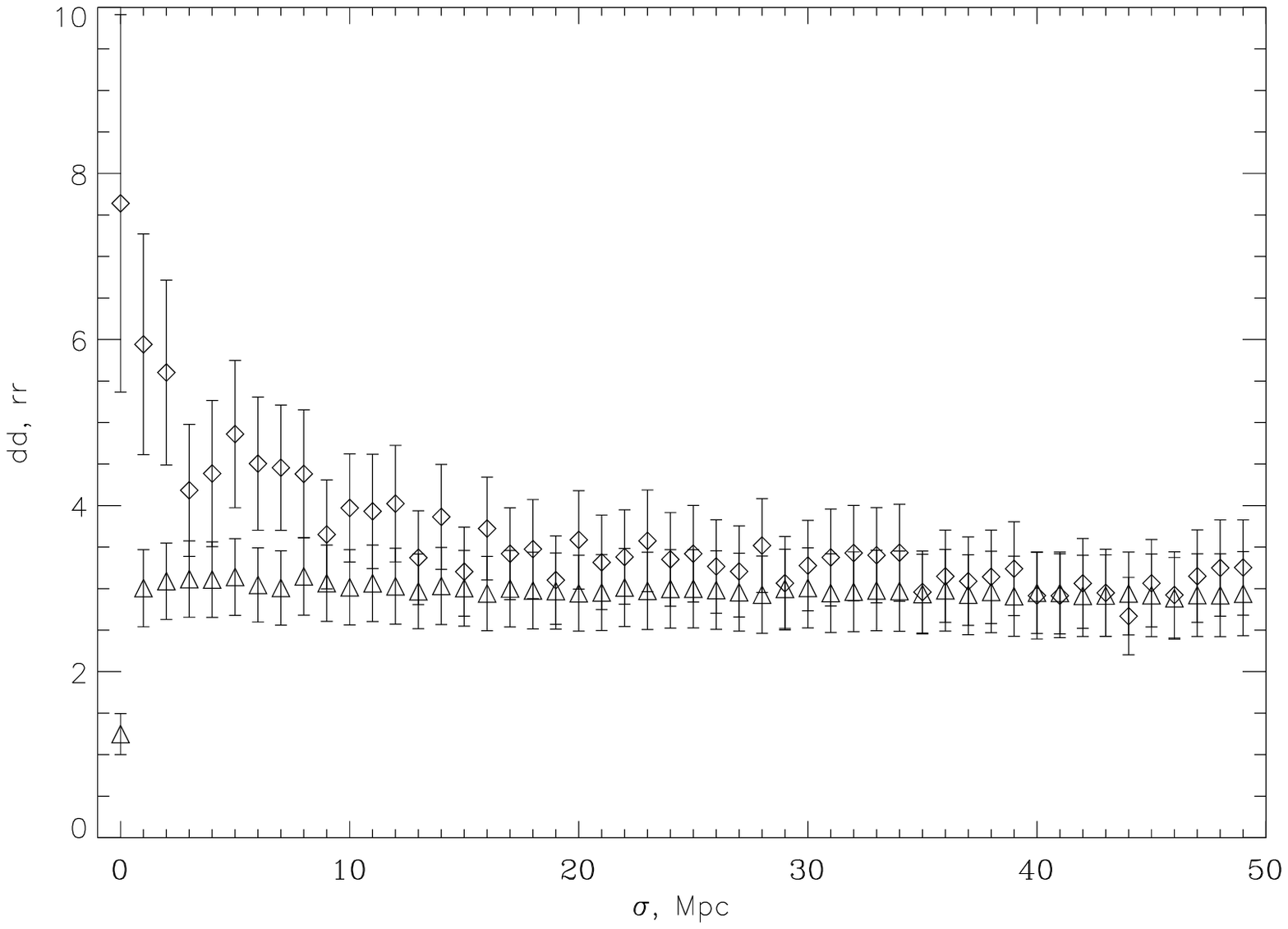,width=8cm}\hfill\epsfig{figure=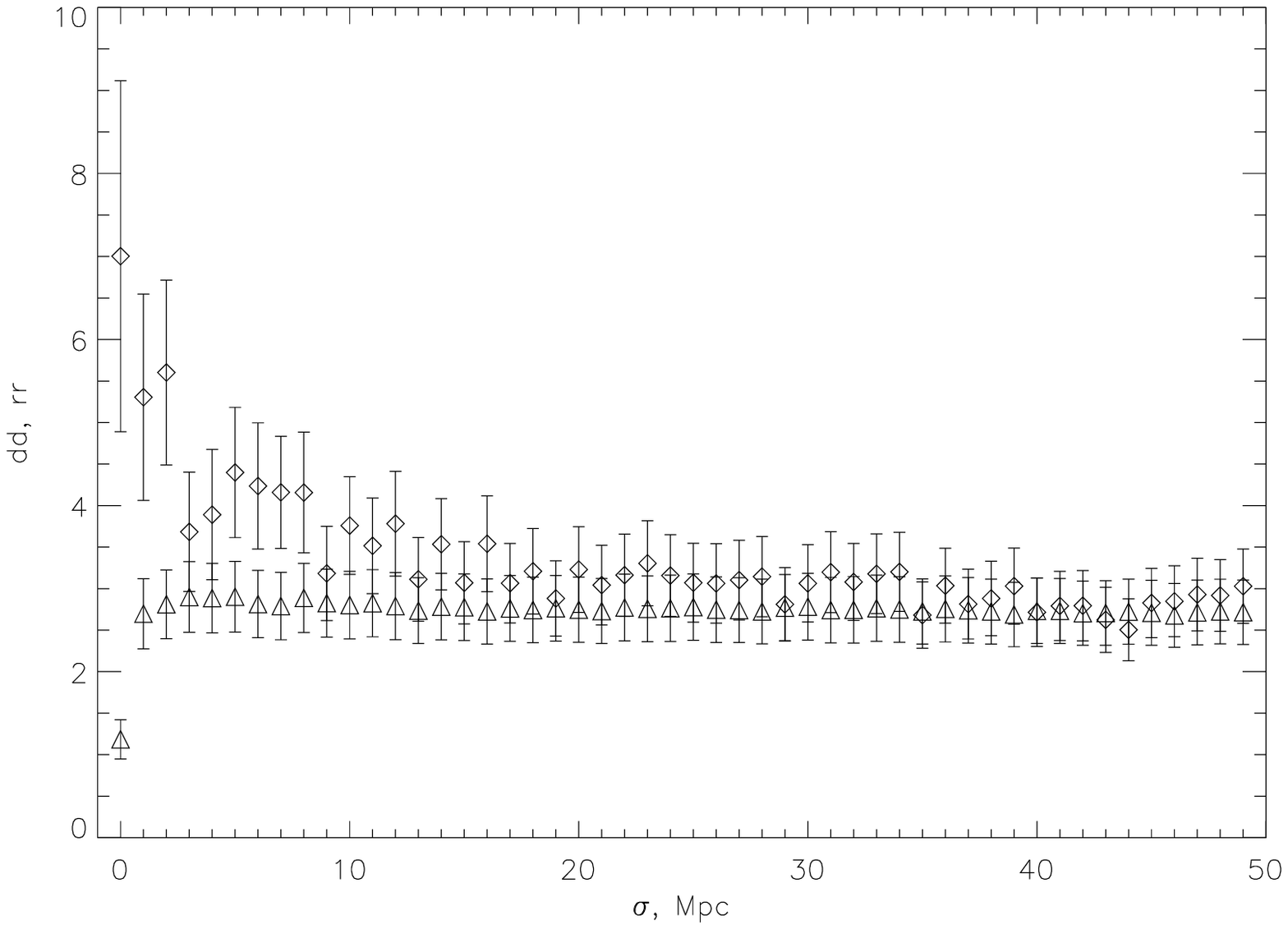,width=8cm}
\caption{Dependence of  normalized neighbour numbers $dd$ (for the initial sample) and $rr$ (for the random sample) on projected distance $\sigma$. Top: \textit{full} sample, bottom: \textit{test}, described in text. Circles denote initial sample, triangles denote random sample.}
 \label{fig_3_4}
\end{figure}

\subsection{Results}\label{sec:3.3}

\indent\indent We determine the distances between quasars using the locally Lorentz frame of the first quasar (see Appendix). This method differs, e.g., from that by \citet{hoyle_2002,daAngela_2005}.

\begin{table*}\label{tab:r-space}
 \centering
 \begin{minipage}{170mm}
  \caption{The parameters of the  real-space 2pCF for different cases described in text and the results of other authors for comparison. Here DR5Q stands for quasar catalogue by Schneider et al. (2007).}
\centering
  \begin{tabular}{|c|c|c|c|c|c|c|c|}
  \hline
dist. range, $h^{-1}$Mpc & $z$ range & mean $z$ & $r_{0}$, h$^{-1}$ Mpc & $\gamma$ & $\chi^{2}/d.o.f.$ & sample & authors\\
  \hline
  &      & & $5.85\pm0.33$ & $1.87\pm0.07$ & 1.054 &         \textit{full}        &                    \\
              &           &      & $5.83\pm0.36$ & $1.87\pm0.08$ & 2.084 & \textit{test} &  \\
\multicolumn{2}{c}{df} &      & $5.83\pm0.36$ & $1.87\pm0.08$ & 2.084 & \textit{test} &  \\
              &           &      & $5.76\pm0.84$ & $1.78\pm0.26$ & --    & \textit{homogeneous} &  \\
$\sigma=1-35$ & $0.8-2.2$ & 1.47 & $6.08\pm0.59$ & $1.88\pm0.11$ & 0.789 & \textit{uniform} & present paper \\
              &           &      & $5.84\pm0.39$ & $1.87\pm0.09$ & 1.231 & \textit{good} &  \\
              &           &      & $5.63\pm0.55$ & $1.88\pm0.12$ & 0.846 & \textit{low reddening} &  \\
              &           &      & $5.73\pm0.57$ & $1.85\pm0.11$ & 1.025 & \textit{high reddening} &  \\
\hline
$r=1-20$ & $0.8-2.1$ & 1.47 & $4.8^{+0.9}_{-1.5}$ & $1.5\pm0.2$ & -- & 2QZ & Porciani et al. (2004) \\
$r=1-10$  & $0.3-2.2$ & 1.4 & $6.0^{+0.5}_{-0.7}$ & $1.45\pm0.27$ & -- & 2QZ & \cite{daAngela_2005}\\
$r=10-40$ & $0.3-2.2$ & 1.4 & $7.25$ & $2.30^{+0.12}_{-0.03}$ & -- & 2QZ & \cite{daAngela_2005} \\
$\sigma=4-150$ & $2.9-5.4$ & -- & $15.2\pm2.7$ & $2.0\pm0.3$ & -- & SDSS DR5Q & \cite{shen_2007} \\
$\sigma=1-130$ & $0.3-2.2$ & 1.27 & $5.45^{+0.35}_{-0.45}$ & $1.90^{+0.4}_{-0.03}$ & -- & SDSS DR5Q & \cite{ross_2009} \\
\hline
\end{tabular}
\end{minipage}
\end{table*}

Our sample covers only a part of the sky; thus the boundary effects can influence the result because the number of neighbours for objects lying close to the boundary is less then for those lying far from it. To estimate this influence we made calculations in two different ways. In the first one we calculated the number of neighbours for all the quasars in the sample. In the second one we took into account the neighbours of the objects lying at least 5 degrees far from the boundary (we called this \textit{test} sample). Fig.~\ref{fig_3_4} shows the surface number density of neighbours $dd=DD/[\pi\Delta\sigma(2\sigma+\Delta\sigma)]$ in the initial sample (circles) and $rr=RR/[\pi \Delta\sigma(2\sigma+\Delta\sigma)]$ in the random one (triangles) correspondingly, where we choose the bin size $\Delta\sigma=1$. Values $RR$ are the mean numbers over 50 realizations of random sample, error bars represent 1$\sigma$ jack-knife errors.

\begin{figure}
\centering
\epsfig{figure=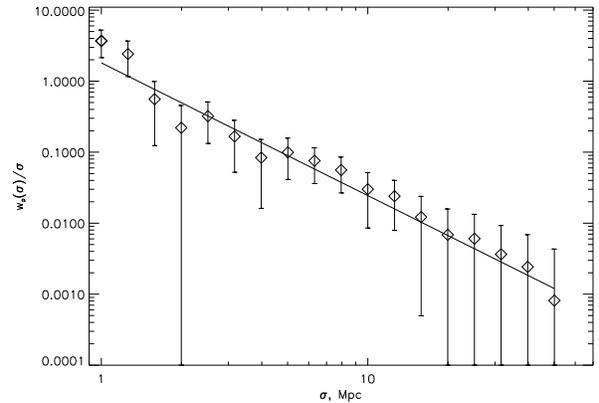,width=8cm}
\caption{The quasar projected 2pCF, $w_{p}$, for \textit{full} sample. The quoted errorbars are jackknifes. The solid line showes the best fit single power-low with $r_{0}=5.85\,h^{-1}$~Mpc and $\gamma=1.87$ over our full range $1<\sigma<35\,h^{-1}$~Mpc.}
 \label{fig_proj_in}
\end{figure}

As can be seen from Fig.~\ref{fig_3_4}, in practice the surface density of neighbours for the random catalogue does not depend on the distance (except for the distance range $\lesssim 1-2$~Mpc), which is expected for a random distribution of objects. The gap at $\sim 1-2$~Mpc can be explained by a deficit of close pairs in the initial catalogue. Note that angular distribution of objects in our random samples is the same as in the initial one (see Sec.~\ref{sec:2.7}). This deficit is a result of the fibre collision effect (Sec.~\ref{sec:2.6}). One can see that the number densities for initial and random samples coinside up to $40-50$~Mpc, thus we use $L=50$~Mpc (or $35\,h^{-1}$~Mpc). And we also put $L_0=1\,h^{-1}$~Mpc because of the fiber collision effect. The quasar projected 2pCF $w_{p}(\sigma)$ divided on $\sigma$ in logarithmic scale for the \textit{full} sample is presented in Fig.~\ref{fig_proj_in} with the the best fit single power-low over the range $1<\sigma<35\,h^{-1}$~Mpc.

\begin{figure}
\centering
\epsfig{figure=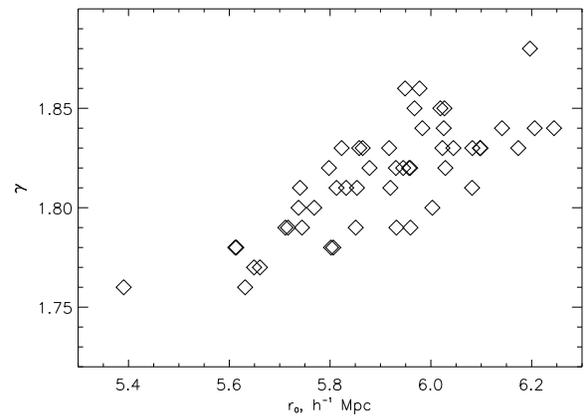,width=8cm}
\caption{Correlation lengthes and slopes of the real-space 2pCF for 50 \textit{homogeneous} samples.}
 \label{fig_pp}
\end{figure}

\begin{figure}
\centering
\epsfig{figure=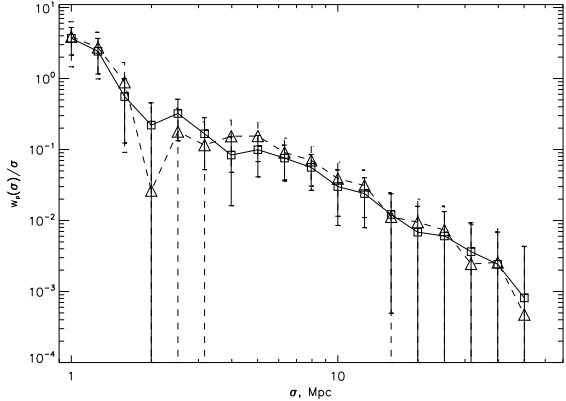,width=8cm}
\caption{The quasar projected 2pCF, $w_{p}$, for \textit{initial} (squares and solid line) and \textit{uniform} (triangles and dashed lines) samples. The quoted errorbars are jackknifes.}
 \label{fig_proj_uni}
\end{figure}

\begin{figure}
\centering
\epsfig{figure=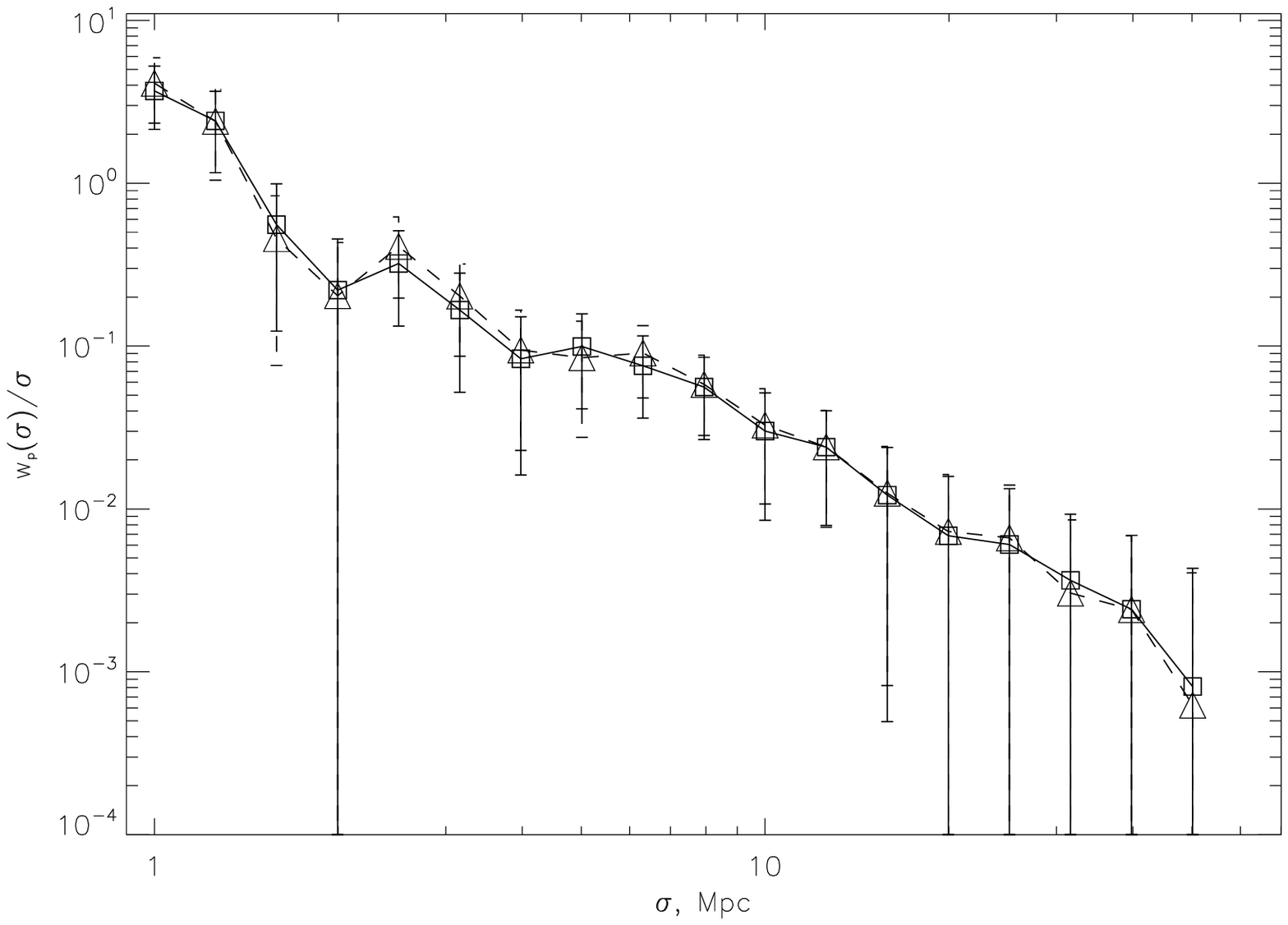,width=8cm}
\caption{The quasar projected 2pCF, $w_{p}$, for \textit{initial} (squares and solid line) and \textit{good} (triangles and dashed lines) samples. The quoted errorbars are jackknifes.}
 \label{fig_proj_good}
\end{figure}

\begin{figure}
\centering
\epsfig{figure=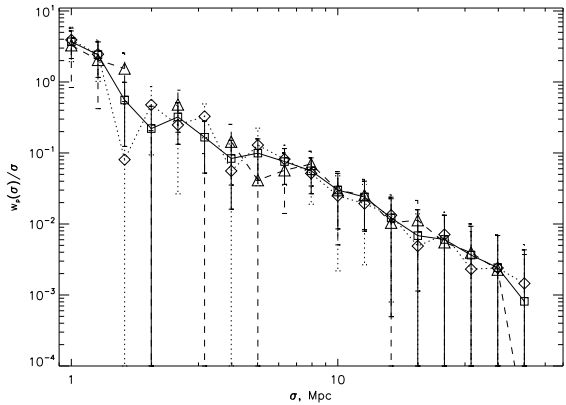,width=8cm}
\caption{The quasar projected 2pCF, $w_{p}$, for \textit{initial} (squares and solid line), \textit{low-reddening} (triangles and dashed lines) and \textit{high-reddening} (diamonds and dotted lines) samples. The quoted errorbars are jackknifes.}
 \label{fig_proj_red}
\end{figure}

The same calculations were provided for 50 test samples (see Sec.~\ref{sec:2.2}). The scattering of the resulting parameters is presented in Fig.~\ref{fig_pp} with the mean values presented in the second row of Table~1, where errors are rms. Results for \textit{homogeneous}, \textit{uniform}, \textit{good}, \textit{low reddening} and \textit{high reddening} samples are presented in rows $3-7$ of this table. For these samples the errors are jackknifes. Comparison of the projected 2pCF for the \textit{initial} sample with \textit{uniform}, \textit{good}, \textit{low reddening} and \textit{high reddening} samples are presented in Figs.~\ref{fig_proj_uni}-\ref{fig_proj_red}.


\subsection{Discussion}\label{sec:3.4}

\indent\indent As one can see from Figs.~\ref{fig_proj_uni}-\ref{fig_proj_red} and Table~1, the projected 2pCFs and parameters of the real-space 2pCF for all samples agree well with each other within the errors that are slightly larger for the samples with smaller number of objects. Hence, firstly, we can neglect boundary effects for our sample, that is expected due to the sample size being several times larger than the distance range we consider. Secondly, we can also neglect all the inhomogeneities of the sample caused by the non-uniformity of the sky coverage as a result of the spectroscopic survey, reddening and the conditions of observations. The last five rows of Table~1 show comparisons of our results with the results of other authors. We can see that our results agree well with early results for the 2QZ catalogue by \citet{porciani_2004} within 1$\sigma$ for $r_{0}$ and 2$\sigma$ for $\gamma$. The best agreement is with the results by \citet{ross_2009} for the 4th edition of SDSS quasar catalogue (compiled from the 5th data release of SDSS by \citet{Schneider_2007}), but note that they used the wider distance range. The 2pCF parameters from \citet{shen_2007} are included to the table only for completeness of known results on the quasars 2pCF in the real-space as they correspond to larger redshifts. Our results cannot be directly compared with that by \citet{daAngela_2005} for 2QZ catalogue, though there is rather a good agreement if we average their values over both distance ranges $1-10$ and $10-40$~$h^{-1}$~Mpc. Note that our redshift range is smaller, but the fraction of quasars contained within the $0.3<z<0.8$ range is small compared with those in the $0.8<z<2.2$ range. Finally, we note that comparison of 2pCF parameters on different distance intervals shows that the power-law form of 2pCF must be replaced by more complicated one for larger distances.

\section[]{Redshift-Space correlation function and infall parameter}\label{sec:4}

\subsection{The technique}\label{sec:4.1}

\indent\indent 2pCF in the redshift-space is not spherically-symmetric because of the effects of velocities mentioned in Sec.~\ref{sec:1}. Here we pay attention mainly to the scales larger than 10~Mpc, where we expect the density contrast to be sufficiently small. In this region the effect of bulk gravitational infall leads to flattening of redshift-space 2pCF contours and to certain enhancement of the monopole part $\xi_{0}(s)$ of this 2pCF, which is obtained by averaging over all angles. On the other hand, in this region we neglect `Finger of God' effect due to random velocities.

The relation of the real-space 2pCF to the redshift-space 2pCF is well known within the linear theory of cosmological perturbations \citep{kaiser_1987,hamilton_1992,matsubara_1996}. The monopole (i.~e. angle averaged) part of 2pCF is proportional to the real-space 2pCF \citep{kaiser_1987} and therefore it also has a power-law form $\xi_{0}(s)=\left(s_{0}/s\right)^{\gamma_s}$ with the same slope $\gamma_s=\gamma$, but with different correlation length:
\begin{equation}\label{eq_beta}
s_{0}=r_{0}\left[
1+\frac{2\beta}{3}+\frac{\beta^{2}}{5}\right]^{1/\gamma_s}.
\end{equation}
\noindent This formula is well-grounded in the region $r > 10$~Mpc, where the linear theory is applicable. Here the parameter $\beta=f(\Omega_{M},z)/b$  appears as some consequence of the large-scale infall onto matter overdensities, $b$ is the bias parameter in Eq.~(\ref{bias1}). The function $f(\Omega_{M},z)$ is expressed by means of the growing mode of the density contrast \citep{peebles_book}. There is a simple analytical approximation \citep{carroll_1992} for this function in the case of the spatially-flat $\Lambda$CDM cosmological model:
\begin{equation}\label{eq:fom}
 f(\Omega_{M},z)\approx\left[\frac{\Omega_{M}(1+z)^{3}}{\Omega_{M}(1+z)^{3}+1-\Omega_{M}}\right]^{4/7}
\end{equation}

Therefore we may proceed in two ways: either (i) by taking the slope $\gamma_s=\gamma$ from the results of Sec.~\ref{sec:3} dealing with fitting of $s_0$ only, or (ii) by independent determination of $\gamma_s,\, s_0$.

Like in Sec.~\ref{sec:3} one can determine the total number of neighbours of the $i$-th quasar from the whole sample in a spherical layer $[s,s+\Delta s]$ as
\[
    DD(s,\Delta s)=\sum_{i}n_{i}\left[1+\frac{4\pi}{\Delta V}\int_{s}^{s+\Delta
    s}\xi(s')s'^{2}ds'\right]\Delta V,
\]
\noindent where $\Delta V=4\pi \Delta(s^2+s\Delta s+(\Delta s)^2/3)$. The similar estimation for random catalogue, which is considered to represent random spatial distribution of objects with no clustering, is
\[
    RR(s,\Delta s)=4\pi \sum_{i} n'_{i}\Delta V.
\]
\noindent Assuming $\sum\limits_{i}n'_{i}\approx\sum\limits_{i}n_{i}$ and taking into account the power-low form of the monopole part of 2pCF $\xi_0(s)=\left(s_{0}/s\right)^{\gamma_s}$, we have the relation for fitting:
\begin{equation}\label{fit}
  \frac{DD(s,\Delta s)}{RR(s,\Delta s)}-1=\frac{s_{0}^{\gamma_s}}{3-\gamma_s}
  \cdot \frac{(s+\Delta s)^{3-\gamma_s}-s^{3-\gamma_s}}{\Delta s\,(s^{2}+s\Delta s+\Delta s^{2}/3)}.
\end{equation}

In order to determine the infall parameter $\beta$ on the way (i) we use Eq.~(\ref{fit}) to find $s_0$, with the slope $\gamma_s=\gamma$ being known from the results of the Sec.~\ref{sec:3}. Then we derive $\beta$ by solving the Eq.~(\ref{eq_beta}) with known $s_0,\,r_0$, and then we estimate bias $b$ for our redshift interval using Eq.~(\ref{eq:fom}). These operations deal with the region $s>10$~Mpc. In addition, mainly for comparison with the results of the other authors, we also considered way (ii) with independent determination of $\gamma$ and $s_0$ for different separation intervals.

Note that the method we use for estimation of the infall parameter is similar to the so-called $J_{3}(s)/J_{3}(r)$-method (see e.g. \citet{ratcliffe_IV_1998}), which uses the volume integrals $J_{3}(x)=\int\limits_{0}^{x}\xi(y)y^{2}dy$. Using of these integrals allows to smooth the influence of small sizes of the samples. Our approach has the same advantage as it includes the 2pCF parameters which are the averaged characteristics of 2pCFs within the whole distance range.

\subsection{Results and discussion}\label{sec:4.2}

\begin{figure}
\centering
\epsfig{figure=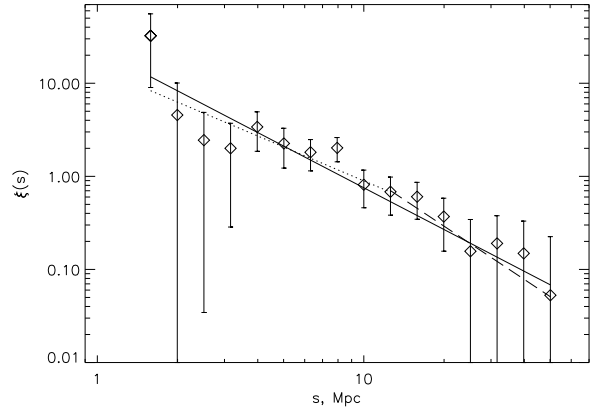,width=8cm}
\caption{The quasar redshift-space 2pCF, $\xi(s)$, for \textit{full} sample. The quoted errorbars are jackknifes. The best fit single power-low over our full range $1-35\,h^{-1}$~Mpc and over ranges $1-10\,h^{-1}$~Mpc and $10-35\,h^{-1}$~Mpc with parameters presented in Table~2 are shown with solid, dotted and dashed lines correspondingly.}
 \label{fig_zsp_in}
\end{figure}

\indent\indent  We estimated the 2pCF parameters within the whole range $s=1-35$~$h^{-1}$~Mpc and separately for ranges $1-10$~$h^{-1}$~Mpc and $10-35$~$h^{-1}$~Mpc. The results are presented in Table~2 and Fig.~13. The errors are calculated using the jackknife technique. We also found the same parameters for distance range $1-25$~$h^{-1}$~Mpc for comparison with the results of \citet{croom_2005} and \citet{ross_2009} presented in the last 3 rows of this table. As one can see our parameters $s_{0}$ and $\gamma_{s}$ within the distance range $1-25$~$h^{-1}$~Mpc agree well with the results of \citet{croom_2005} and \citet{ross_2009} for 2QZ catalogue and the 4th edition of the SDSS quasar catalogue \citep{Schneider_2007} correspondingly. And for distance range $1-10$~$h^{-1}$~Mpc our results agree with \citet{croom_2005} within 2$\sigma$ for the slope and 1$\sigma$ for the correlation length.

\begin{table*}\label{tab:s-space}
 \centering
 \begin{minipage}{170mm}
  \caption{The parameters of  the redshift-space  2pCF and the results of other authors for comparision. Here DR5Q stands for quasar catalogue by Schneider et al. (2007).}
\centering
  \begin{tabular}{|c|c|c|c|c|c|c|c|}
  \hline
dist. range, $h^{-1}$Mpc & $z$ range & mean $z$ & $s_{0}$, h$^{-1}$ Mpc & $\gamma_{s}$ & $\chi^{2}/d.o.f.$ & sample & authors\\
  \hline
& & & $5.83\pm0.47$ & $1.49\pm0.12$ & 1.440 & \textit{initial} & \\
& & & $5.95\pm0.86$ & $1.44\pm0.21$ & -- & \textit{homogeneous} & \\
& & & $5.33\pm0.51$ & $1.35\pm0.08$ & 1.473 & \textit{uniform} & \\
$s=1-35$ & $0.8-2.2$ & 1.47 & $5.76\pm0.54$ & $1.46\pm0.12$ & 1.363 & \textit{good} & present paper \\
 &  &  & $5.45\pm0.96$ & $1.48\pm0.24$ & 0.779 & \textit{low reddening} &  \\
& & & $4.56\pm0.55$ & $1.24\pm0.11$ & 1.892 & \textit{high reddening} & \\
\hline
$s=1-25$ & & & $5.97\pm0.51$ & $1.37\pm0.13$ & 1.060 &  &  \\
$s=1-10$ & $0.8-2.2$ & 1.47 & $6.43\pm0.63$ & $1.21\pm0.24$ & 0.456 & \textit{initial} & present paper \\
$s=10-35$ & & & $7.37\pm0.81$ & $1.90\pm0.24$ & 1.636 & &  \\
\hline
$s=1-25$ & $0.3-2.2$ & 1.35 & $5.48^{+0.42}_{-0.48}$ & $1.20\pm0.10$ & -- & 2QZ & \cite{croom_2005}\\
$s=1-10$ & $0.3-2.2$ & 1.35 & $3.88^{+0.43}_{-0.53}$ & $0.86^{+0.16}_{-0.17}$ & -- & 2QZ & \cite{croom_2005}\\
$s=1-25$ & $0.3-2.2$ & 1.27 & $5.95\pm0.45$ & $1.16^{+0.11}_{-0.16}$ & -- & SDSS DR5Q & \cite{ross_2009} \\
\hline
\end{tabular}
\end{minipage}
\end{table*}

\begin{table*}\label{tab:b-and-beta}
 \centering
 \begin{minipage}{170mm}
  \caption{The values of infall parameter $\beta$ and bias parameter $b_{Q}$ and the results of other authors for comparison. Here DR5Q stands for quasar catalogue by Schneider et al. (2007),   DR4phot stands for a sample of photometrically classified quasars from the   4th Data Release of SDSS. The numbers in brackets correspond to methods proposed by \protect\citet{hoyle_2002} -- [1], \protect\citet{daAngela_2005} -- [2], \protect\citet{croom_2005} -- [3], and \protect\citet{porciani_2004} -- [4]; see explanations in the text.}
\centering
  \begin{tabular}{|c|c|c|c|c|c|c|}
  \hline
mean $z$ & $b_{Q}$ & $\beta$ & method & sample & authors\\
  \hline
0.55 & $1.4\pm0.2$   & $0.55\pm0.10$ & mean of [1],[2] & 2SLAQ+2QZ+SDSS DR5 & \cite{Mountrichas_2009} \\
1.27 & $2.06\pm0.03$ & $0.43$        & [3] & SDSS DR5Q & \cite{ross_2009} \\
1.35 & $2.02\pm0.07$ & $0.44\pm0.02$ & [3] & 2QZ & \cite{croom_2005}\\
1.40 & $1.50\pm0.20$ & $0.60^{+0.14}_{-0.11}$ & [2] & 2QZ+2SLAQ & \cite{daAngela_2008} \\
1.40 & $2.84^{+1.49}_{-0.57}$ & $0.32^{+0.09}_{-0.11}$ & $\xi(s)/\xi(r)$ & 2QZ & \cite{daAngela_2005} \\
1.40 & -- & $0.50^{+0.13}_{-0.15}$ & [2] & 2QZ & \cite{daAngela_2005} \\
1.40 & $2.41\pm0.08$ & -- & [3] & SDSS DR4phot & \cite{myers_2006_1} \\
1.47 & $2.42^{+0.20}_{-0.21}$ & --    & [4] & 2QZ   & \cite{porciani_2004} \\
\hline
 & $1.44\pm0.22$ & $0.63\pm0.10$ &  & \textit{initial} &  \\
 & $1.37\pm0.22$ & $0.64\pm0.10$ &  & \textit{homogeneous} &  \\
1.47 & $1.50\pm0.37$ & $0.61\pm0.15$ & decribed in text & \textit{uniform} & present paper \\
 & $1.35\pm0.23$ & $0.67\pm0.12$ &  & \textit{good} &  \\
 & $1.30\pm0.44$ & $0.70\pm0.24$ &  & \textit{low reddening} &  \\
 & $1.54\pm0.51$ & $0.67\pm0.12$ &  & \textit{high reddening} &  \\
\hline
\end{tabular}
\end{minipage}
\end{table*}

Comparing the  parameters for different distance ranges we can conclude that on the large scales the redshift-space 2pCF is steeper than on the small scales while the correlation lengthes do not differ significantly within 1$\sigma$. This difference in the slope of the redshift-space 2pCF of quasars on different scales was pointed out earlier  \citep{croom_2005}. It is important to note that the results for the slope on $10-35\,h^{-1}$~Mpc interval fairly well agree with the results for $\gamma$ in the real space, as it must be.

\begin{figure}
\centering
\epsfig{figure=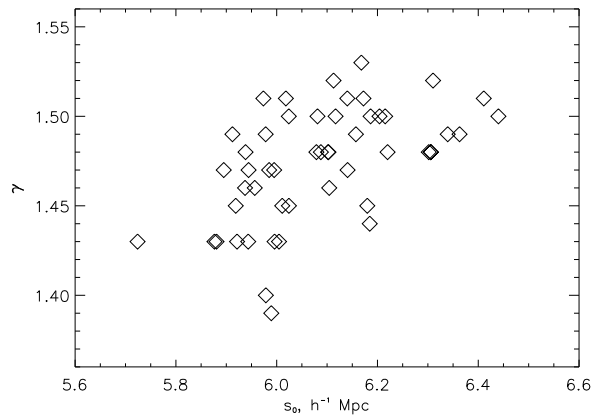,width=8cm}
\caption{Correlation lengthes and slopes of the redshift-space 2pCF for 50 \textit{homogeneous} samples.}
 \label{fig_sp}
\end{figure}
\begin{figure}
\centering
\epsfig{figure=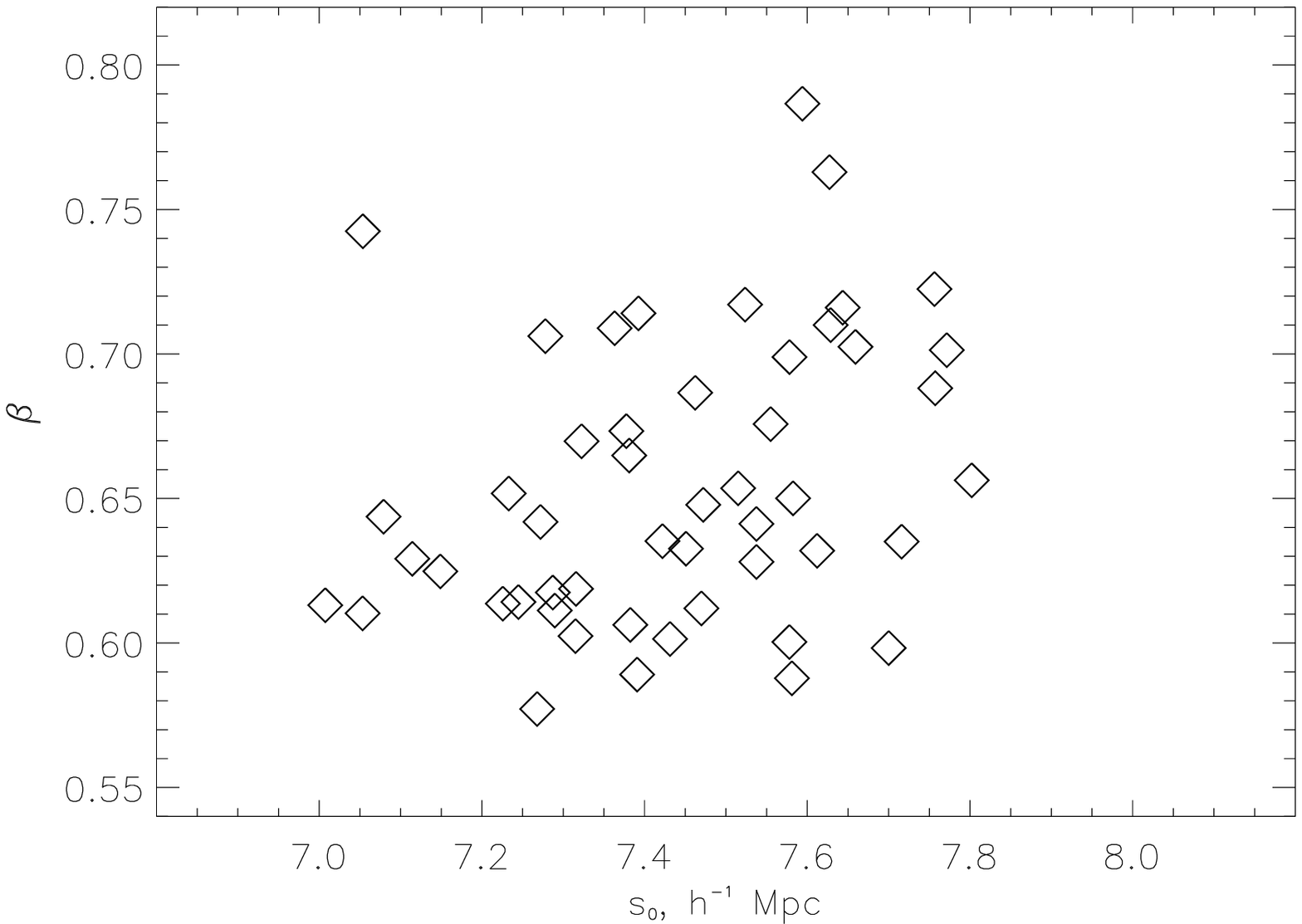,width=8cm}
\caption{Correlation lengthes (for fixed slope value) and infall parameters for 50 \textit{homogeneous} samples.}
 \label{fig_bs}
\end{figure}

As discussed in Section~4.1, we have used the correlation length $r_{0}$ and the slope $\gamma$ of the real-space 2pCF obtained in Section~\ref{sec:3}. We have then obtained the redshift-space correlation length $s_0=7.27\pm0.37\,h^{-1}$~Mpc for the fixed $\gamma$ on the scales $>10\,h^{-1}$~Mpc and we have found $\beta$ and $b$ using Eqs.~(\ref{eq_beta}) and (\ref{eq:fom}). The results are given in Table~3 that contains the obtained values of infall and bias parameters and results of other authors with indication of the method, used for their estimation. One of them (denoted as [1]), containing simultaneous determination of $\beta$ and cosmological model, was proposed by \citet{hoyle_2002} and includes fitting of the observed $\xi(\sigma,\pi)$ by theoretical model from \citet{peebles_book} (see also \citet{ratcliffe_IV_1998}) with the relation for infall velocity from \citet{hale-sutton_1990}. The improvement of this method proposed in \citet{daAngela_2005} (denoted as [2]) allows to break the degeneracy between $\Omega_{M}$ and $\beta$ and is based in the consideration of the quasar clustering evolution. Another approach proposed in \citet{croom_2005} (denoted as [3]) suggests the solution of the quadratic equation in $b_{Q}$ with the shape of $\xi(r)$ governed by the underlying dark matter distribution described by the analytic equation from \citet{hamilton_1991,hamilton_1995}. In \citet{daAngela_2008} and \citet{ross_2009} the last approach was applied with theoretical dark matter power spectrum from \citet{smith_2003}. One more method which was used in \citet{porciani_2004} (denoted as [4]) includes using the relation between projected 2pCFs of the quasars and that of the matter. As distinct from a number of authors (aimed, in particular, at estimations of the dark matter halo masses) we concentrate to the comparison  with the linear theory; i.e. our results on $b_{Q}$ are completely based on the interval $r>10\,h^{-1}$~Mpc, where there is no problems with the linear theory.

For test we also provided the same calculations for redshift-space 2pCF, $\beta$ and $b$ for \textit{homogeneous}, \textit{uniform}, \textit{good}, \textit{low reddening} and \textit{high reddening} samples. The scattering of the resulting redshift-space 2pCF parameters and infall parameters with the corresponding values of $s_{0}$ for 50 \textit{homogeneous} samples are presented in Figs.~\ref{fig_sp} and \ref{fig_bs}. Their mean values are presented in Tables~2 and 3 together with results for other samples. For all our samples errors of the parameters are jackknifes, except the \textit{homogeneous} where errors are rms. All these results agrees well (within 1.3$\sigma$) with the results for initial sample.

\section{Random velocities dispersion}\label{sec:6}

\subsection{Technique}\label{sec:6.1}

\indent\indent In this Section we  discuss the `Finger of God' effect as a tool to estimate the random component of the peculiar velocity $v$ of quasars with respect to the CMB frame, i.e. the dispersion $\langle v^{2}\rangle$. We neglect virialization and assume that the random velocities do not correlate with mass density variations (contrary to \citet{kaiser_1987} effect); such assumption is quite reasonable in case of quasars. If the quasars in a pair can be considered as independent, the line-of-sight pairwise velocity dispersion is $\langle w^{2}\rangle= 2 \,\langle v^{2}\rangle$, where one-dimensional pairwise velocity $w$ is the difference of the peculiar velocities (above the Hubble flow) $v_{1}$, $v_{2}$ of quasars in a pair: $w=v_1-v_2$. The effect of random velocities typically dominates at smaller separations, where the gravitational infall is less pronounced. Here, we neglect the effect of the gravitational infall.

There are two major obstacles. Firstly, as it was mentioned above, the SDSS data are incomplete in the region of linear comoving projected separations $\sigma<L_0=2$~Mpc. Nevertheless one can try to use an information from this region by testing different distributions. Here we assume that the selection effects due to the fibre collisions do not distort considerably the distribution of quasar pairs from our sample with projected comoving separations $\sigma<L_{0}=2$~Mpc and the redshift $0.8<z<2.2$. Another obstacle is the poor knowledge of the 2pCF form in this region. This is the main point of this Section: to study the model dependence of the results  for  $\langle v^{2}\rangle^{1/2}$. In order to obtain  $\langle v^{2}\rangle^{1/2}$, we need 2pCF for all distances. Therefore, we used the power-law form of the real-space 2pCF within the comoving distance range $2<r<50$~Mpc with the parameters defined in Sec.~\ref{sec:3} and assumed a rough estimate for $r \le L_{0}$:
\begin{equation}
\label{dcf_small} \xi(r) = \left\{ {\begin{array}{l}
 (r_{0}/r)^\gamma,\quad\quad\quad\quad\quad\quad r >L_{0}, \\
 \eta(r_{0}/L_{0})^\gamma(L_{0}/r)^{\gamma_1 },\quad r \le L_{0}. \\
 \end{array}} \right.
\end{equation}
According to Sec.~\ref{sec:3} we  used $\gamma = 1.85$, $r_{0}=5.8\,h^{-1}$~Mpc and $\gamma_{1}$, $\eta$ will be varied ($\gamma_{1}<3$ to avoid divergence for $r\to 0$). In case of $\eta=1$ 2pCF is represented by a continuous function. On account of \citet{hennawi_06} results, the rough choice $\gamma_{1}=0$, $\eta=10$ may be considered as plausible. However, even rather wild ($\eta=100$) choices do not lead to a drastic differences of results. Namely, we saw that the results are rather stable (taking into account errors) under different choices of $\gamma_{1}$, $\eta$.

Formula (\ref{dcf_small}) allows to estimate the number of pairs with separation $r$. Then we restricted ourselves with the region $\sigma<L_{0}$ and use relation
\[
\psi(\Pi)=2\pi\int\limits_0^{L_0}d\sigma\,\sigma\,\xi\left(\sqrt{\Pi^2+\sigma^2
}\right)
\]
\noindent to derive  distribution of pair numbers (minus the random background) in line-of-sight separations $\Pi$.

We considered two kinds of the pairwise velocity distribution: Gaussian and exponential \citep{ratcliffe_IV_1998, hoyle_2002,hawkins_2003}.  The pairwise velocity $w$ contributes to the change of the comoving line-of-sight distance $\Pi$ between the  quasars in a pair as follows: $\Pi\to|\Pi+\delta|$, where $\delta=w/[H_0 K(z)]$ with $K(z)=[\Omega_{M}(1+z)^3+1-\Omega_{M}]^{1/2}/(1+z)$. This coefficient grows from 0.85 to 0.97 within the interval $0.8<z<2.2$, hence we approximate it as $K(z)\approx K(1.5)\approx 0.9$.

First we assumed the exponential pairwise velocity distribution \citep{ratcliffe_IV_1998}. Then we have probability density $\Phi(\delta)$ of random variable $\delta$ as follows 
\begin{equation}\label{eq:zh3}
\Phi(\delta)=\frac{\alpha}{2}\exp\left(-\alpha|\delta|\right),\quad \alpha =\frac{\sqrt{2} K H_0}{\langle w^{2}\rangle^{1/2}}.
\end{equation}
In case of Gaussian distribution of velocities we have
\begin{equation}\label{gauss}
\Phi(\delta)=\frac{1}{\sqrt{2\pi\sigma^2}} \exp\left(-\frac{\delta^2}{2\sigma^2}\right),\quad \sigma = \frac{{\langle w^{2}\rangle}^{1/2}}{K H_0}.
\end{equation}
The  convolution of $\psi$ with $\Phi$ yields the distribution of pair numbers  as a function of $\Pi$. We used this to fit the observed pair numbers in different bins with the  projected separation $\sigma<L_{0}=2$~Mpc after randomized background subtraction. The randomized background is calculated as described in Sec.~\ref{sec:2.7}.

\subsection{Results and discussion}\label{sec:6.2}
\begin{figure}
\centering
\epsfig{figure=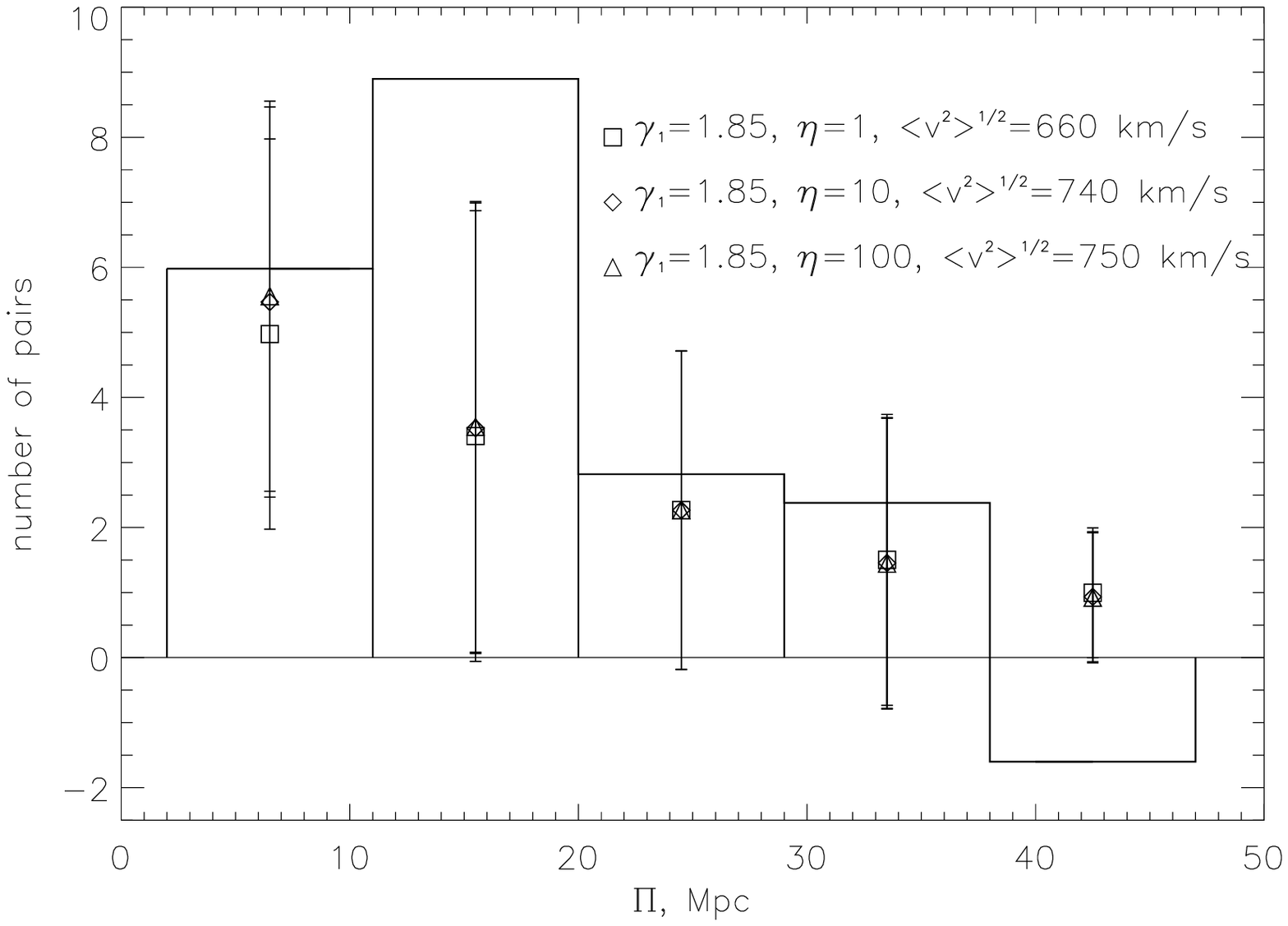,width=8cm}
\epsfig{figure=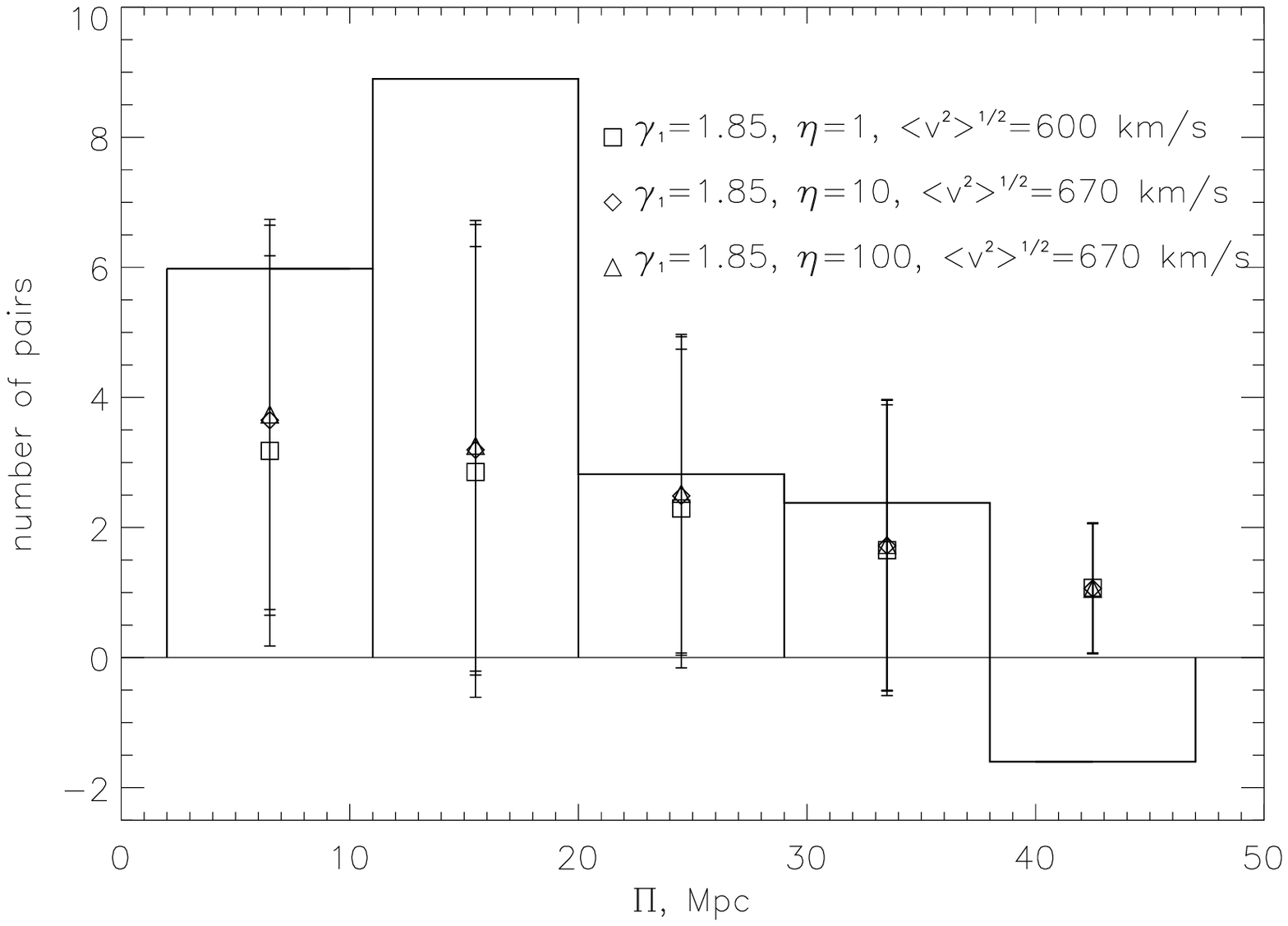,width=8cm}
\caption{Line-of-sight distance distribution of pair numbers in case of Binning 2 with fitted distributions for different values of $\gamma_1$, $\eta$, $\langle v^2\rangle^{1/2}$ for exponential (top) and Gaussian (bottom) distributions. The rectangles show the background subtracted pair numbers, the symbols with errors show the results of fitting.}
 \label{fig_vel_eg1}
\end{figure}

\begin{figure}
\centering
\epsfig{figure=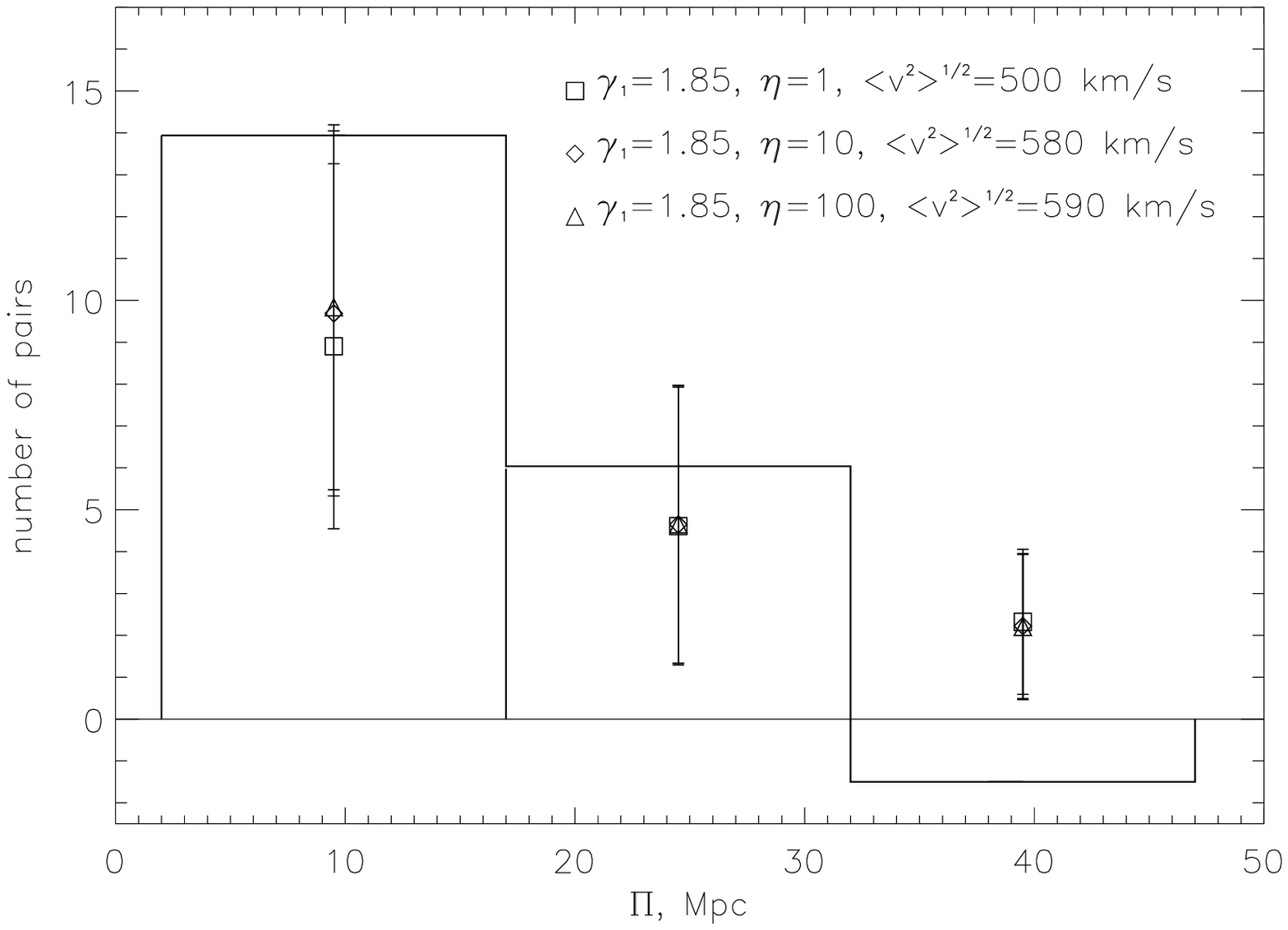,width=8cm}\hfill
\epsfig{figure=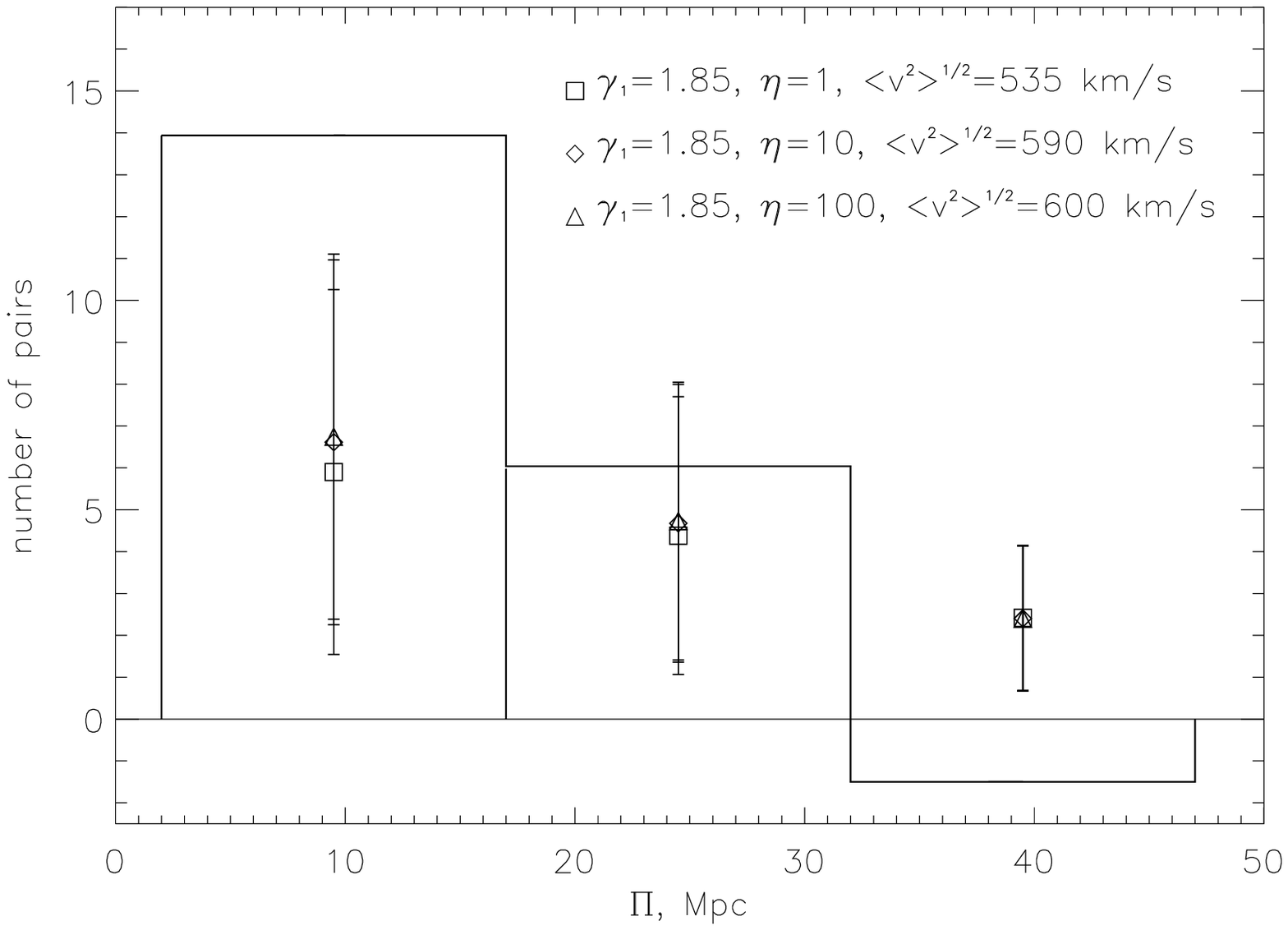,width=8cm}
\caption{The same as in {\protect Fig.~\ref{fig_vel_eg1}} in case of Binning 2.}
 \label{fig_vel_eg2}
\end{figure}

\indent\indent Because  the number of pairs is not large, there is a considerable variation of results for different binnings, so we do not consider the numerical estimates as very reliable. However, we may argue that the model dependence on $\gamma_1$ and $\eta$ parameters is rather insignificant. For illustration, we present the results (Table~4) of fitting the pair numbers (after randomized background subtraction) on account of (\ref{eq:zh3}) and (Table~5) on account of (\ref{gauss}) for two different binnings having minimal jackknife dispersion estimates for $\langle v^2\rangle^{1/2}=\langle w^2/2\rangle^{1/2}$. Binning 1 involves numbers of pairs with line-of-sight separations $Pi$ from five intervals $[2,11]$, $[11,20]$, $[20,29]$, $[29,38]$, $[38,47]$~(Mpc). Binning 2 deals with separations from three intervals $[2,17]$, $[17,32]$, $[32,47]$~(Mpc). The bar charts  in Fig.~\ref{fig_vel_eg1} and \ref{fig_vel_eg2} we show corresponding numbers after the average background subtraction. Here we prefer to avoid separations less than 2~Mpc. Also for illustration in Fig.~\ref{fig_vel_eg1} and \ref{fig_vel_eg2} we present the results of fitting for $\gamma_1=1.85$ and different values of $\eta$. As one can see from the results, the values of $\langle v^2\rangle^{1/2}$ do not vary considerably under rather different (even unrealistic) choices of unknown parameters $\eta$, $\gamma_1$: the variations are of the order of the jackknife dispersion estimate.


\begin{table*}
 \begin{minipage}{80mm}\label{tab:zh2}
\centering
  \caption{The random velocity dispersion ${\langle v^{2}\rangle}^{1/2}$ (km/s) for exponential distribution. Upper values correspond to Binning 1, lower - to Binning 2.}
  \begin{tabular}{|c|c|c|c|}
\hline
& $\gamma_1 = 0$& $\gamma _1 = 1.85$& $\gamma _1 = 2.9$ \\
\hline
$\eta = 1$& 620$\pm $140& 660$\pm $150& 740$\pm $160 \\
 & $460\pm180 $& $500\pm $180& 580$\pm $190 \\
\hline
$\eta = 10$& 720$\pm $160& 740$\pm $170& 750$\pm $170 \\
 & 550$\pm $190& 580$\pm $190& 590$\pm $190 \\
\hline
$\eta = 100$& 750$\pm $165& 750$\pm $165& 750$\pm $165 \\
 & 590$\pm $190& 590$\pm $190& 590$\pm $190 \\
\hline
\end{tabular}
\end{minipage}\hfill
 \begin{minipage}{80mm}\label{tab:zh2G}
\centering
  \caption{The random velocity dispersion ${\langle v^{2}\rangle}^{1/2}$ (km/s) for Gaussian distribution. Upper values correspond to Binning 1, lower - to Binning 2.}
  \begin{tabular}{|c|c|c|c|}
\hline
& $\gamma_1 = 0$& $\gamma _1 = 1.85$& $\gamma _1 = 2.9$ \\
\hline
$\eta = 1$& 580$\pm $90& 600$\pm $90& 670$\pm $90 \\
 & $490\pm 120 $& $535\pm $125& 600$\pm $110 \\
\hline
$\eta = 10$& 650$\pm $100& 670$\pm $100& 690$\pm $80 \\
 & 560$\pm $125& 590$\pm $130 & 600$\pm $120 \\
\hline
$\eta = 100$& 670$\pm $90& 670$\pm $90& 690$\pm $95 \\
 & 580$\pm $130& 600$\pm $100& 600$\pm $120 \\
\hline
\end{tabular}
\end{minipage}
\end{table*}

In fact  the values $\langle v^2\rangle^{1/2}$  in Tables~5,\,6 do not represent real quasar velocity dispersion, but a superposition of two components: $\langle v^2\rangle = v_{err}^2 + \langle \bar{v}^2\rangle$, where $\langle\bar{v}^2\rangle^{1/2}$ is the true  velocity dispersion and $v_{err}$ is due to the redshift measurement errors. The last value corresponds to intrinsic emission lines shifts in quasars pointed out by \citet{croom_2005}, as it is an estimation from emission-line or cross-correlation technique of redshift measurement caused by the impossibility of precise determination of the line centre due to its large width. For detailed investigation of this effect see \citet{richards_2002,shen_2007}. The redshift measurement errors
from the SDSS data base for our sample  is $\Delta z=0.0019 $, which corresponds to $v _{err}\sim c\Delta z/(1+\bar{z})=230$~km/s. In this case we would obtain too large value of $\langle\bar{v}^2\rangle$.

Once work on the first version of the text had been finished, the paper by \citet{hewett_10} was appeared with the error estimate $v_{err}=600$~km/s for SDSS data. In this case $z$-errors dominate in ${\langle v^2\rangle}^{1/2}$. If we take the results from Table~5, which yield smaller jackknife dispersion estimate for $\langle{v}^2\rangle^{1/2}$, than the residual upper estimate of $\langle\bar{v}^2\rangle^{1/2}\le (690^2-600^2)^{1/2}=340$~km/s seems to be more adequate.

The values of the pairwise velocity dispersion $\langle w^2\rangle= 2 \langle v^2\rangle$ corresponding to Tables~4,\,~5 are comparable with that mentioned in previous studies of quasar clustering. \citet{outram_2001} used the value 400~km/s obtained from the Hubble Volume Simulations; \citet{daAngela_2005} fixed the value 800~km/s and found that this value gives an adequate result for $s<10$~$h^{-1}$~Mpc region and noted that it is dominated by the rms pairwise redshift error $600$~km/s; the same value 800~km/s was used by \citet{daAngela_2008} and the close value 690~km/s was used by \citet{croom_2005}. The last value was chosen as a mean one of the range 630-750~km/s; this is a quadrature superposition of (i) the real pairwise velocity dispersion recalculated with redshift estimated from the galaxies pairwise velocity dispersion 500~km/s \citep{hawkins_2003}, (ii) redshift measurement error obtained from repeat observations, and (iii) velocity error  due to intrinsic emission lines shifts in QSOs \citep{richards_2002}. Only \citet{Mountrichas_2009} treating the velocity dispersion as a free parameter estimated its value for a sample of quasars and luminous red galaxies (LRG) as 620~km/s and 727~km/s (two values are the result of two different approaches) and noted that they are consistent with the value 728~km/s which is a quadrature superposition of the values 800~km/s for quasars and 300~km/s for LRG from previous studies.

\section{Conclusions}\label{sec:7}

\indent\indent We have  performed a detailed analysis of 2pCF using the quasar sample from the SDSS DR7 data for objects with $0.8<z<2.2$. Our main conclusions are as follows.
\begin{enumerate}
 \item Using  our approach for estimation of the real-space 2pCF $\xi(r)$ of quasars we confirm  that it is fitted well with the power-low model in the distance range $1<r<35$~$h^{-1}$~Mpc with the  correlation length of $r_{0}=5.85\pm0.33$~$h^{-1}$~Mpc and the power-low slope  $\gamma=1.87\pm0.07$. The results are in a good agreement with  the results of the other authors on the same distance interval. However, this comparison suggests that the power-low model seems to be not valid on larger intervals.
 \item The redshift-space 2pCF of  quasars has a break and it may be approximated well with the double power-low model with $s_{0}=6.43\pm0.63$~$h^{-1}$~Mpc and $\gamma=1.21\pm0.24$ for $1<s<10$~$h^{-1}$~Mpc and $s_{0}=7.37\pm0.81$~$h^{-1}$~Mpc and $\gamma=1.90\pm0.24$ on scales $10<s<35$~$h^{-1}$~Mpc with the mean values $5.83\pm0.47$~$h^{-1}$~Mpc and $\gamma=1.49\pm0.12$ for both distance ranges together.
 \item  For distances $s>10\,h^{-1}$~Mpc the parameter describing the large-scale infall to density inhomogeneities is $\beta=0.63\pm0.10$ with the linear bias to be $b=1.44\pm0.22$ that marginally (within 2$\sigma$) agrees with the linear theory of cosmological perturbations.
 \item We point out that the value of the quasars  velocity dispersion varies almost within errors for a wide range of model parameters that determine 2pCF for  $r<2$~Mpc.
 \end{enumerate}

We also performed investigations of our sample and showed that we can neglect boundary effects and all inhomogeneities of the sample caused by non-uniformity of the sky coverage by the spectroscopic survey, reddening and conditions of observations.

It is worth to note, that the values of $\langle w^2 \rangle^{1/2}$ from Tables~4,\,5 have  been calculated on account of SDSS DR7 data on pairs with the  projected separations $\sigma<2$~Mpc, which are not complete and the number of  pairs involved in treatment of the velocity dispersion is not too large. In  view of these circumstances we consider our results on  $\langle w^2\rangle^{1/2}$ rather as motivation to further studies to observe larger sample of $\sim 1$~Mpc quasar pairs with accurate redshift measurements.

\section*{Acknowledgments}

\indent\indent This work has been  partially supported by Swiss National Science Foundation  (SCOPES grant No 128040).

The authors are thankful  to the Sloan Digital Sky Survey team. Funding for the SDSS has been provided by the Alfred P. Sloan Foundation, the Participating Institutions, the National Aeronautics and Space Administration, the National Science Foundation, the US Department of Energy, the Japanese Monbukagakusho, and the Max Planck Society. The SDSS Web site is http://www.sdss.org.

The authors are also thankful to the anounimous referee for helpful comments and suggestions.

\appendix

\section[]{Distances determination}

\indent\indent We consider the spatially flat Universe with Friedmann-Robertson-Walker (FRW) metric
\begin{equation}
ds^{2}=c^{2}dt^{2}-a^{2}(t)\left[d\chi^{2}+\chi^{2}\left(d\delta^{2}+\cos^{2}\delta d\alpha^{2}\right) \right],
\end{equation}
\noindent where $\alpha$ and $\delta$  are right ascention and declination respectively, $a(t_{0})=1$ at the present epoch $t_{0}$. In case of the flat $\Lambda$CDM cosmological model ($\Omega_{M}+\Omega_{\Lambda}=1$) for an object with redshift $z$ we have
\[
\chi(z)=\frac{c}{H_{0}}\int_{0}^{z}\frac{d\zeta}{\sqrt{\Omega_{\Lambda}+\Omega_{M}(1+\zeta)^{3}}}=t(z).
\]
\noindent The distances in the redshift-space are defined as if the objects were at rest with respect to the CMB. In order to define the distance
$r_{qN}$ from a quasar with coordinates $\{t_q,\chi_q,\alpha_q,\delta_q\}$ to a neighbouring object with coordinates $\{t_N,\chi_N,\alpha_N,\delta_N\}$  we proceed as follows \citep{ivzh_10}.

(a) Consider the Fermi frame (Synge 1960; Misner et al. 1973) of the quasar $q$ with the fiducial trajectory $x^{\mu}(t)=\{t,\chi_q,\alpha_q,\delta_q\}$ that may be interpreted as the trajectory of an observer, which is at rest with respect to the CMB.

(b) Introduce the hypersurface of constant time in corresponding normal Fermi coordinates with the origin $\{t_q,\chi_q,\alpha_q,\delta_q\}$. In our case this is the space-like hypersurface $t=t_q$, which is orthogonal to the fiducial trajectory at the origin.

(c) Find the invariant (proper) distance $R_{qN}$ from the origin to the intersection of the neighbour world line with the hypersurface; this is uniquely determined along the space-like geodesic connecting two points. In a general case we must suppose that the objects are not too far from each other to avoid the conjugate points.

In our case $R_{qN}$ is easily determined along the geodesic on the hypersurface $t=t_q$ between the points $\{t_q,\chi_q,\alpha_q,\delta_q\}$ and $\{t_q,\chi_N,\alpha_N,\delta_N\}$; then the comoving distance $r_{qN}$ is:
\begin{equation}
 \label{r_total}
 r_{qN}=(1+z_q)R_{qN}=\sqrt{\chi_q^{2}+\chi_N^2-2\chi_q\chi_N\cos(\psi_{qN})},
\end{equation}
\noindent where $\psi_{qN}$ is the angle between objects $q,N$ on the sky. The comoving linear projected distance from $q$ to $N$ is
\begin{equation}
 \label{r_perp} \sigma=\chi_{N}\sin(\psi_{qN}),
\end{equation}
and the comoving line-of-sight distance from $q$ to $N$ is
\begin{equation}
 \label{r_par} \Pi=\chi_{N}\cos(\psi_{qN})-\chi_{q},
\end{equation}

The above formulae determine distances in the redshift-space. Transition to the real-space involves the velocities of quasars with respect to cosmological background. However, in case of the projected distances the effect of velocities is negligible and these distances obtained from observations may be used to derive parameters of the real-space 2pCF.

\bsp
\label{lastpage}
\end{document}